\documentclass[twocolumn]{aastex701}
\usepackage{amsmath}

\begin{document}

\title{Irregularly Sampled Time Series Interpolation for Binary Evolution Simulations\\ Using Dynamic Time Warping}

\author[0000-0001-7273-0211]{Ugur\,Demir}
\affiliation{Electrical and Computer Engineering, Northwestern University, 2145 Sheridan Road, Evanston, IL 60208, USA}
\affiliation{Center for Interdisciplinary Exploration and Research in Astrophysics (CIERA), Northwestern University, 1800 Sherman Ave, Evanston, IL 60201, USA}
\affiliation{NSF-Simons AI Institute for the Sky (SkAI),172 E. Chestnut St., Chicago, IL 60611, USA}
\email{ugur.demir@northwestern.edu}

\author[0000-0003-1749-6295]{Philipp\,M.\,Srivastava}
\affiliation{Electrical and Computer Engineering, Northwestern University, 2145 Sheridan Road, Evanston, IL 60208, USA}
\affiliation{Center for Interdisciplinary Exploration and Research in Astrophysics (CIERA), Northwestern University, 1800 Sherman Ave, Evanston, IL 60201, USA}
\affiliation{NSF-Simons AI Institute for the Sky (SkAI),172 E. Chestnut St., Chicago, IL 60611, USA}
\email{fakeemail2@google.com}

\author[0000-0003-4554-0070]{Aggelos\,Katsaggelos}
\affiliation{Electrical and Computer Engineering, Northwestern University, 2145 Sheridan Road, Evanston, IL 60208, USA}
\affiliation{Center for Interdisciplinary Exploration and Research in Astrophysics (CIERA), Northwestern University, 1800 Sherman Ave, Evanston, IL 60201, USA}
\affiliation{NSF-Simons AI Institute for the Sky (SkAI),172 E. Chestnut St., Chicago, IL 60611, USA}
\email{fakeemail2@google.com}

\author[0000-0001-9236-5469]{Vicky\,Kalogera}
\affiliation{Center for Interdisciplinary Exploration and Research in Astrophysics (CIERA), Northwestern University, 1800 Sherman Ave, Evanston, IL 60201, USA}
\affiliation{NSF-Simons AI Institute for the Sky (SkAI),172 E. Chestnut St., Chicago, IL 60611, USA}
\affiliation{Department of Physics and Astronomy, Northwestern University, 2145 Sheridan Road, Evanston, IL 60208, USA}
\email{fakeemail2@google.com}

\author[0000-0003-2090-7446]{Santiago\, L.\,Tapia}
\affiliation{Electrical and Computer Engineering, Northwestern University, 2145 Sheridan Road, Evanston, IL 60208, USA}
\affiliation{Center for Interdisciplinary Exploration and Research in Astrophysics (CIERA), Northwestern University, 1800 Sherman Ave, Evanston, IL 60201, USA}
\affiliation{NSF-Simons AI Institute for the Sky (SkAI),172 E. Chestnut St., Chicago, IL 60611, USA}
\email{fakeemail2@google.com}

\author[0000-0001-9135-6421]{Manuel\,Ballester}
\affiliation{Electrical and Computer Engineering, Northwestern University, 2145 Sheridan Road, Evanston, IL 60208, USA}
\affiliation{Center for Interdisciplinary Exploration and Research in Astrophysics (CIERA), Northwestern University, 1800 Sherman Ave, Evanston, IL 60201, USA}
\affiliation{NSF-Simons AI Institute for the Sky (SkAI),172 E. Chestnut St., Chicago, IL 60611, USA}
\email{fakeemail2@google.com}

\author[0000-0002-2677-8019]{Shamal\,Lalvani}
\affiliation{Electrical and Computer Engineering, Northwestern University, 2145 Sheridan Road, Evanston, IL 60208, USA}
\email{fakeemail2@google.com}
 
\author[0009-0003-1268-5529]{Patrick\,Koller}
\affiliation{Electrical and Computer Engineering, Northwestern University, 2145 Sheridan Road, Evanston, IL 60208, USA}
\affiliation{Center for Interdisciplinary Exploration and Research in Astrophysics (CIERA), Northwestern University, 1800 Sherman Ave, Evanston, IL 60201, USA}
\affiliation{NSF-Simons AI Institute for the Sky (SkAI),172 E. Chestnut St., Chicago, IL 60611, USA}
\email{fakeemail2@google.com}

\author[0000-0001-5261-3923]{Jeff\, J.\,Andrews}
\affiliation{Department of Physics, University of Florida, 2001 Museum Rd, Gainesville, FL 32611, USA}
\affiliation{Institute for Fundamental Theory, 2001 Museum Rd, Gainesville, FL 32611, USA}
\email{fakeemail2@google.com}

\author[0000-0001-6692-6410]{Seth\,Gossage}
\affiliation{Center for Interdisciplinary Exploration and Research in Astrophysics (CIERA), Northwestern University, 1800 Sherman Ave, Evanston, IL 60201, USA}
\affiliation{NSF-Simons AI Institute for the Sky (SkAI),172 E. Chestnut St., Chicago, IL 60611, USA}
\email{fakeemail2@google.com}

\author[0000-0002-6842-3021]{Max\,M.\,Briel}
\affiliation{Département d’Astronomie, Université de Genève, Chemin Pegasi 51, CH-1290 Versoix, Switzerland}
\affiliation{Gravitational Wave Science Center (GWSC), Université de Genève, CH1211 Geneva, Switzerland}
\email{fakeemail2@google.com}

\author[0000-0003-0420-2067]{Elizabeth\,Teng}
\affiliation{Center for Interdisciplinary Exploration and Research in Astrophysics (CIERA), Northwestern University, 1800 Sherman Ave, Evanston, IL 60201, USA}
\affiliation{NSF-Simons AI Institute for the Sky (SkAI),172 E. Chestnut St., Chicago, IL 60611, USA}
\affiliation{Department of Physics and Astronomy, Northwestern University, 2145 Sheridan Road, Evanston, IL 60208, USA}
\email{fakeemail2@google.com}

\begin{abstract}

Binary stellar evolution simulations are computationally expensive. Stellar population synthesis relies on these detailed evolution models at a fundamental level. Producing thousands of such models requires hundreds of CPU hours, but stellar track interpolation provides one approach to significantly reduce this computational cost. Although single-star track interpolation is straightforward, stellar interactions in binary systems introduce significant complexity to binary evolution, making traditional single-track interpolation methods inapplicable. Binary tracks present fundamentally different challenges compared to single stars, which possess relatively straightforward evolutionary phases identifiable through distinct physical properties. Binary systems are complicated by mutual interactions that can dramatically alter evolutionary trajectories and introduce discontinuities difficult to capture through standard interpolation. In this work, we introduce a novel approach for track alignment and iterative track averaging based on Dynamic Time Warping to address misalignments between neighboring tracks. Our method computes a single shared warping path across all physical parameters simultaneously, placing them on a consistent temporal grid that preserves the causal relationships between parameters. We demonstrate that this joint-alignment strategy maintains key physical relationships such as the Stefan-Boltzmann law in the interpolated tracks. Our comprehensive evaluation across multiple binary configurations demonstrates that proper temporal alignment is crucial for track interpolation methods. The proposed method consistently outperforms existing approaches and enables the efficient generation of more accurate binary population samples for astrophysical studies.

\end{abstract}



%%%%%%%%%%%%%%%%%%%%%%%%%%%%%%%%%%%%%%%%%%%%%%%%%%%%%%%%%%%%%%%%%%%%%%%%%%%
\section{Introduction}
The characterization of binary star evolution represents a fundamental challenge in modern astrophysics with significant implications for stellar population studies. Binary stellar evolution, which simulates the behavior and properties of binary systems over their entire lifetimes, is modeled through detailed numerical integration of coupled differential equations \citep{2011ApJS..192....3P, 2013ApJS..208....4P,2015ApJS..220...15P,2018ApJS..234...34P,paxton_modules_2019,2023ApJS..265...15J}. While these physics-based simulations provide high-fidelity representations of binary evolution, simulating stellar populations or conducting parameter studies can require hundreds of CPU hours for population synthesis \citep{COSMIC_2020ApJ...898...71B}.
%While these physics-based simulations provide high-fidelity representations of binary evolution, simulating stellar populations or conducting parameter studies can require hundreds of CPU hours for population synthesis \citep{COSMIC_2020ApJ...898...71B}. Consequently, to implement computationally feasible binary population synthesis, codes opt to use less accurate but computationally cheaper representations of binary evolution. 
%Historically, this has been achieved in primarily two ways.

Binary population synthesis has been addressed through two primary approaches. The first approach employs fitting formula to calculate the evolution of individual stars \citep[][calculated with full stellar structure and evolution codes]{1995MNRAS.274..964P, 1997MNRAS.291..732T, 2000MNRAS.315..543H}, complemented by prescriptions that model the effects of binary interactions \citep[e.g.,][]{2001A&A...365..491N, 2002MNRAS.329..897H, 2008ApJS..174..223B, 2012A&A...546A..70T, 2018MNRAS.480.2011G, 2018MNRAS.481.1908K, 2019MNRAS.485..889S, 2020ApJ...898...71B, 2021arXiv210910352T}. Similarly, the SEVN code \citep{Spera_2019, Iorio_2023} performs population synthesis by interpolating single-star evolutionary tracks from pre-computed grids, using phase-specific coordinates to achieve structure-preserving alignment, with binary interactions handled through analytic and semi-analytic prescriptions. The second approach includes population synthesis methods that consider full stellar structure to calculate binary evolutions, such as {\tt BPASS} \citep{2017PASA...34...58E, 2018MNRAS.479...75S}, which models one binary component, and most recently {\tt POSYDON} \citep[][v1 and v2, respectively]{fragos_posydon_2023, posydon_andrews}, which models both binary components. 

The ability to rapidly generate stellar evolution tracks with arbitrary initial conditions, without requiring full numerical integration for each new configuration, would transform the production of large-scale stellar evolution model grids. This capability would directly address current computational bottlenecks that restrict thorough parameter space exploration in large population synthesis studies.

Interpolating binary evolutionary tracks using pre-simulated grids presents a promising approach for computational acceleration but introduces significant technical challenges. The difficulty in applying traditional interpolation methods to binary systems stems from how binary interactions are modeled. Single-star evolution follows well-understood physical processes that enable clear identification of evolutionary phases \citep{dotter_mesa_2016}. Binary systems, however, involve mass transfer between stars, which is treated approximately in one-dimensional stellar evolution models. While we can identify when mass transfer begins (when a star expands beyond its Roche lobe) and ends (when it contracts back), the actual mass transfer process is not precisely defined. Accurate modeling requires computationally expensive multi-dimensional hydrodynamical simulations, which remain an active research area \citep{2025A&A...702A..61R, 2024ApJ...975..130D}. Current binary evolution codes use various approximate prescriptions for mass transfer rates, creating model-dependent variations that prevent identification of universal evolutionary markers needed for traditional interpolation techniques.

Stellar evolution codes, such as MESA \citep{2013ApJS..208....4P}, model system evolution over timescales of billions of years, adaptively sampling time intervals to concentrate computational resources on critical evolutionary phases where system parameters change rapidly. This adaptive approach produces tracks with variable sequence lengths and irregular temporal sampling across different initial conditions. However, the primary challenge for interpolation is not just the irregular sampling itself, but rather the abrupt morphological changes introduced by binary interactions such as mass transfer events. These interaction-driven changes create discontinuities in evolutionary tracks that complicate alignment between neighboring systems.

\cite{Srivastava_2025} proposed a data compression algorithm that identifies a constant number of representative points from each track to enable averaging-based interpolation for arbitrary initial conditions. Our analysis demonstrates that proper track alignment represents a critical step that significantly impacts prediction accuracy. We show that improved alignment methodologies can substantially enhance interpolation performance across the parameter space.

In this work, we present an improved approach for aligning tracks to accurately interpolate binary star system parameters throughout their evolutionary history. Our methodology leverages pre-simulated binary grids as a memory bank for K-Nearest Neighbor \citep{knn} retrieval and employs Dynamic Time Warping (DTW) \citep{dtw} to align tracks while preserving morphological features introduced by binary interactions. By addressing the fundamental challenges posed by interaction-driven morphological changes through advanced alignment techniques, our approach maintains high physical accuracy while reducing computational time by several orders of magnitude, thereby enabling the efficient generation of substantially larger binary population samples for statistical astrophysical studies.

%%%%%%%%%%%%%%%%%%%%%%%%%%%%%%%%%%%%%%%%%%%%%%%%%%%%%%%%%%%%%%%%%%%%%%%%%%%
%%%%%%%%%%%%%%%%%%%%%%%%%%%%%%%%%%%%%%%%%%%%%%%%%%
\begin{figure*}
%\plotone{figs/grid_introv3.pdf}
\includegraphics[width = \textwidth]{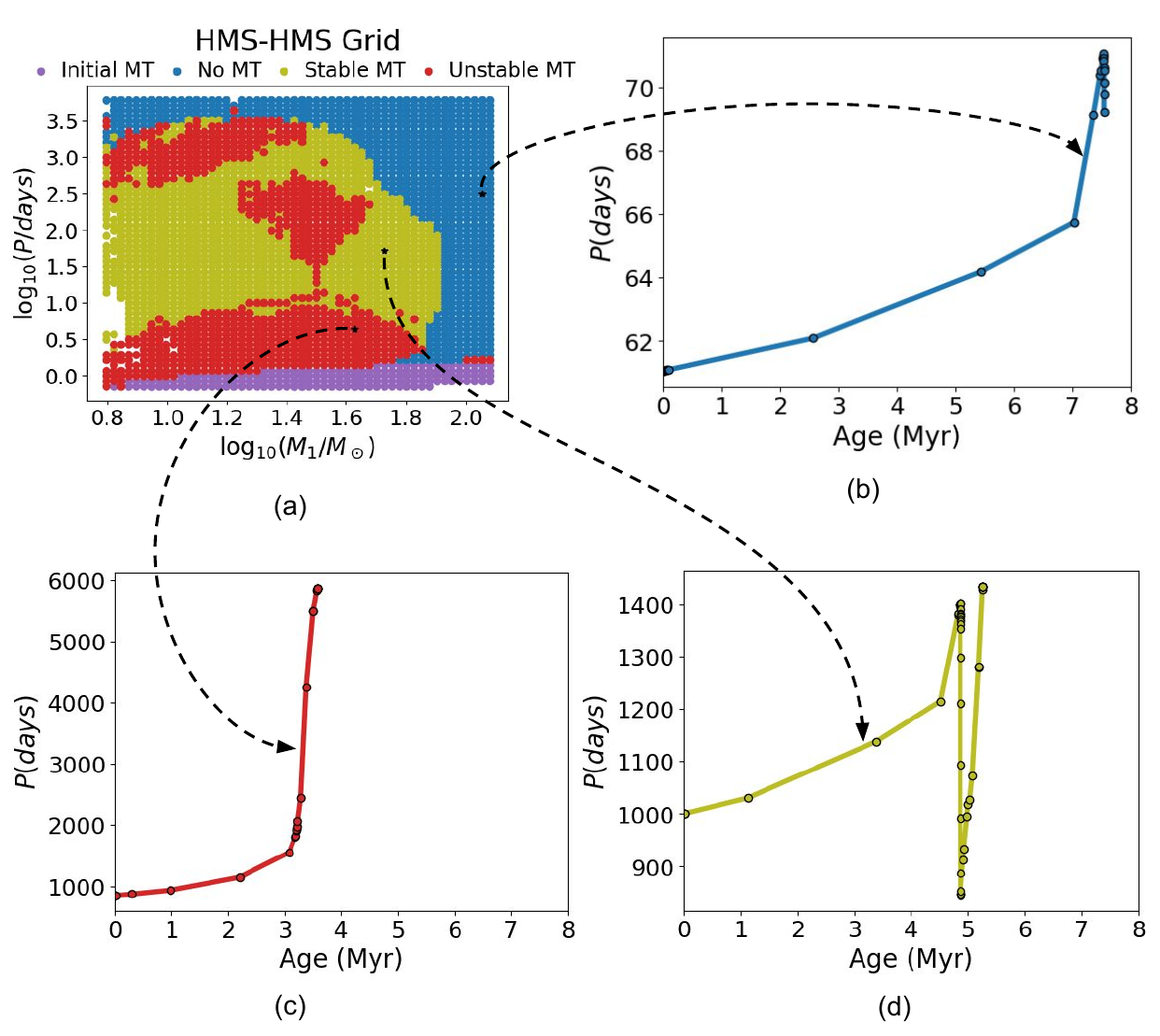}
\caption{(a) A slice from the 3D HMS-HMS grid, displaying primary star mass ($M_1$) on the x-axis and orbital period ($P$) on the y-axis. Both axes are in logarithmic space, resulting in uniform point separation. The secondary star mass dimension ($M_2$) is fixed at $M_2/M_1 = 0.7$ for visualization clarity. Each colored point corresponds to a simulated initial value configuration, with color coding indicating the evolutionary mass transfer class. Uncolored regions represent parameter combinations that were not simulated and are excluded from the dataset. (b), (c) and (d) illustrate example evolutionary tracks for No Mass Transfer, Unstable Mass Transfer, and Stable Mass Transfer classes, respectively. \label{fig:grid_intro}}
\end{figure*}
%%%%%%%%%%%%%%%%%%%%%%%%%%%%%%%%%%%%%%%%%%%%%%%%%%

\section{Background} \label{sec:background}
This study utilizes binary star evolution grids from the POSYDON project \cite{fragos_posydon_2023}, focusing on three distinct evolutionary configurations: Hydrogen Main Sequence-Hydrogen Main Sequence (HMS-HMS), Compact Object-Hydrogen Main Sequence  (CO-HMS), and Compact Object-Helium Main Sequence (CO-HeMS).

Binary star systems within each grid are uniquely characterized by their initial conditions, which consist of the primary star mass, the secondary star mass, and the orbital period. These parameters serve as coordinates in 3D grids. Each grid stores complete evolutionary tracks corresponding to these initial conditions.

For clarity and consistency throughout this paper, we employ the following parameter notation:

\begin{itemize}
    \item $M_1$: Primary star mass (expressed in solar masses $M_\odot$).
    \item $M_2$: Secondary star mass (expressed in solar masses $M_\odot$).
    \item $P$: Orbital period (expressed in days).
    \item $\dot{M}_{transfer}$: The mass transfer rate is represented logarithmically. This parameter quantifies the mass exchange rate between binary components, remaining constant during non-interacting evolutionary phases.
    \item $T_{eff,1}$: Effective temperature of the primary star measured in Kelvin. The values are represented in log scale.
    \item $T_{eff,2}$: Effective temperature of the secondary star measured in Kelvin. The values are represented in log scale.
    \item $L_1$: Luminosity of the primary star measured in solar luminosity ($L_\odot$). The values are represented in log scale.
    \item $L_2$: Luminosity of the secondary star measured in solar luminosity ($L_\odot$). The values are represented in log scale.
    \item $R_1$: The radius of the primary star measured in solar radii ($R_\odot$). The values are represented in log scale.
    \item $R_2$: The radius of the secondary star measured in solar radii ($R_\odot$). The values are represented in log scale.

\end{itemize}

The HMS-HMS grid contains all the parameters listed for every binary star track in the dataset. CO-HMS and CO-HeMS grids do not contain secondary star parameters for luminosity ($L_2$),  effective temperature ($T_{eff,2}$) and the radius ($R_2$).

%%%%%%%%%%%%%%%%%%%%%%%%%%%%%%%%%%%%%%%%%%%%%%%%%%
\subsection{Problem Definition}

In this section, we mathematically define binary star track interpolation. Let us denote a set of initial condition grid points as
\begin{equation}
G = \{i_1, i_2, i_3, \ldots, i_N\},    
\end{equation}
where each grid point $i_j = (M_1^j, M_2^j, P^j)$ represents the $j$-th initial configuration characterized by primary star mass, secondary star mass, and orbital period, respectively.

For each initial condition $i_j \in G$, a corresponding evolutionary track $H_j$ exists, representing the temporal evolution of the binary system parameters. Each track can be expressed as a multivariate time series
\begin{equation}
H_j = \{(a^j_t, \mathbf{v}^j_t) \mid t = 1, 2, \ldots, L_j\},
\end{equation}
where $a^j_t$ denotes the $t$-th time point in the $j$-th track, $\mathbf{v}^j_t \in \mathbb{R}^D$ represents the vector of $D$ physical parameters at that time point, and $L_j$ is the total number of time points in the $j$-th track. The full parameter vector contains $D=10$ components: $\mathbf{v}^j_t = (M_1, M_2, P, \dot{M}_{\rm transfer}, T_{{\rm eff},1}, T_{{\rm eff},2}, L_1, L_2, R_1, R_2)^j_t$ for the HMS-HMS grid, and $D=7$ for the CO-HMS and CO-HeMS grids which lack secondary star parameters for luminosity ($L_2$), effective temperature ($T_{{\rm eff},2}$), and radius ($R_2$) because the secondary (compact object) is represented by a point mass in these grids.
 
An important aspect of our method is the selection of parameters to be used for alignment, that is, a subset $\mathcal{A} \subseteq \{1, \ldots, D\}$ of parameter indices used to compute the warping path during Dynamic Time Warping (DTW) alignment (Section~\ref{sec:dtw}). The resulting warping path is then applied to all $D$ parameters simultaneously, to maintain the temporal consistency across all physical quantities within the same binary system. This joint alignment is critical for preserving physical relationships between coupled parameters such as luminosity, radius, and effective temperature. In the simplest case, all parameters are used for alignment ($\mathcal{A} = \{1, \ldots, D\}$), but subsets can also be chosen to emphasize parameters with the most informative morphological features. For example, in the CO-HMS grid, using only the orbital period and luminosity for alignment yields better interpolation accuracy than using the full parameter set, as discussed in Section~\ref{sec:alignment_selection}.

%-------------------------
%For each initial condition $i_j \in G$, a corresponding evolutionary track $H_j$ exists, representing the temporal evolution of the binary system parameters. Each track can be expressed as a time series 

%\begin{equation}
%H_j = \{(a_t^j, v_t^j) \ | \ t = 1,2,\ldots,\mathcal{L}_j\},    
%\end{equation}
%where $a_t^j$ denotes the $t$-th time point in the $j$-th track, $v_t^j$ represents a corresponding physical parameter and $\mathcal{L}_j$ is the total number of time points in the $j$-th track. The track $H_j$ can represent one of the physical parameters from $ \{M_1, M_2, P, \dot{M}_{transfer}, T_{eff,1}, T_{eff,2}, L_1, L_2, R_1, R_2\}$. Since our proposed method operates on these parameters independently, the methodology applies to each parameter separately. For simplicity, we omit explicit parameter indexing in the notation throughout this paper.

%-------------------------

Each evolutionary track within the binary system grid is systematically classified according to its mass transfer characteristics: No Mass Transfer (NMT), Stable Mass Transfer (SMT), Unstable Mass Transfer (UMT), and Initial Mass Transfer (IMT). Tracks belonging to the same class show significant similarities in their morphological profiles and evolutionary trajectories. Figure \ref{fig:grid_intro} shows how the classes are distributed in an example grid slice.

We address the interpolation problem by generating a predictive evolutionary track $H_*$ for a binary system with initial parameters $i_* = (M_1^*, M_2^*, P^*)$ that are not explicitly represented in the existing grid $G$, aiming to capture the physical evolution. Our method aims to find a mapping $f: i_* \mapsto H_*$ from the data provided by the POSYDON grid simulations. Figure \ref{fig:grid_intro} illustrates a slice ($M_1$, $M_2$, $P$) for a fixed $M_2/M_1$ from the HMS-HMS grid with its initial parameter space and representative simulated evolutionary tracks for specific initial values. The sampling of initial values differs across the grids. For CO-HMS and CO-HeMS grids, the parameters $M_1$, $M_2$, and $P$ are uniformly varied in $\log_{10}$ space, creating regular three-dimensional grids. In contrast, the HMS-HMS grid uses a different sampling strategy: while $M_1$ and $P$ are varied in $\log_{10}$ space, the secondary mass is determined by the mass ratio $M_2/M_1$ rather than being independently varied. This parameterization results in a grid that exclusively contains points where $M_1 > M_2$, ensuring the primary star is always more massive than the secondary.

%%%%%%%%%%%%%%%%%%%%%%%%%%%%%%%%%%%%%%%%%%%%%%%%%%%%%%%%%%%%%%%%%%%%%%%%%%%

\section{Method}
\begin{figure*}
%\plotone{figs/archv6.pdf}
\includegraphics[width = \textwidth]{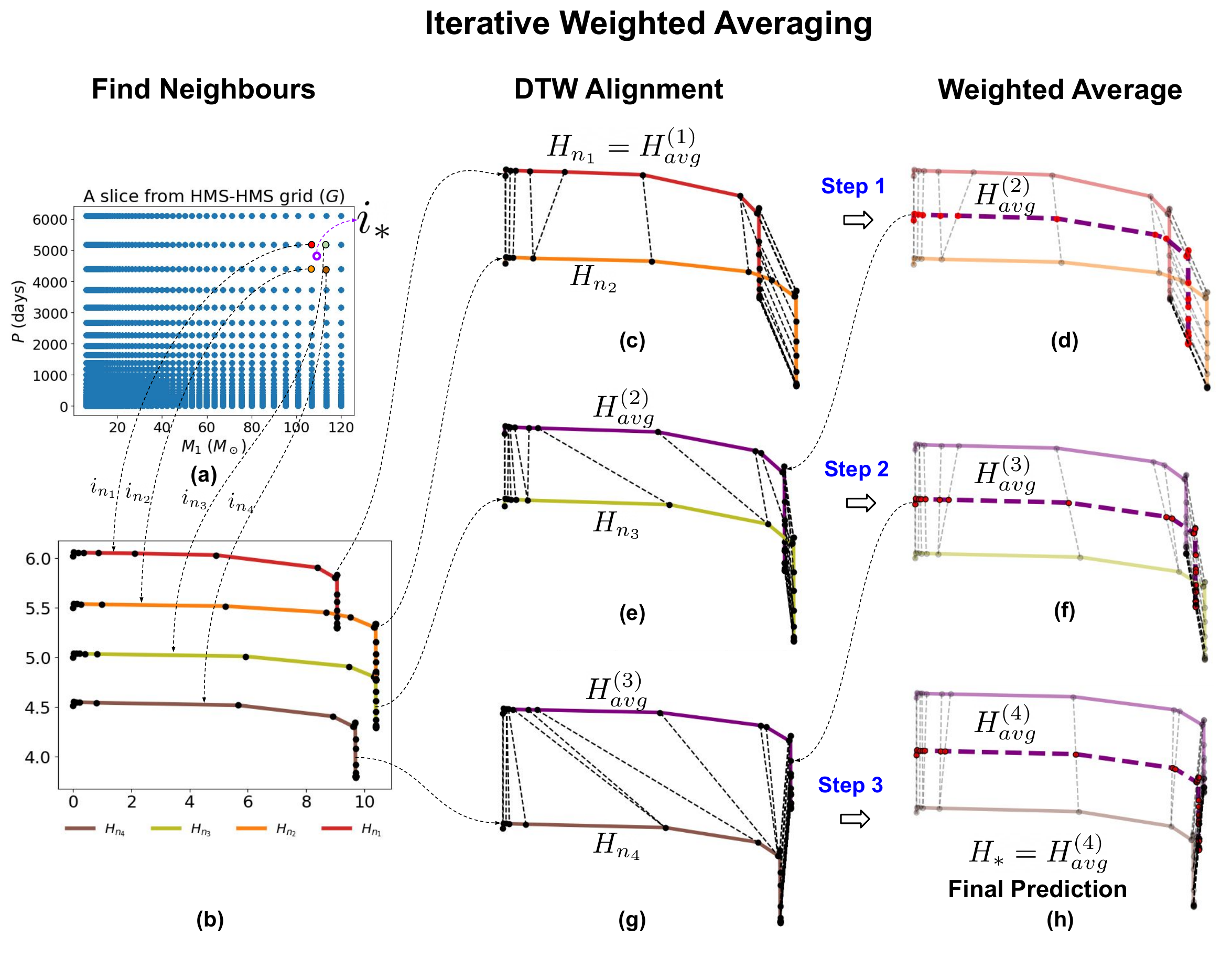}
\caption{Overview of the proposed DTW-based iterative interpolation method. For a given target initial condition $i_*$ shown in (a), we identify the $K$ nearest neighbors (here $K=4$) from the grid $G$: $i_{n_1}$, $i_{n_2}$, $i_{n_3}$, and $i_{n_4}$. The grid shown in (a) is plotted without a $\log$ transformation. We retrieve their corresponding evolutionary tracks $H_{n_1}$, $H_{n_2}$, $H_{n_3}$, and $H_{n_4}$ shown in (b). The interpolation proceeds through iterative alignment and averaging. First, tracks $H_{n_1}$ and $H_{n_2}$ are aligned using DTW as shown in (c), then averaged to produce $H_{\text{avg}}^{(2)}$ (d). Next, $H_{n_3}$ is aligned with $H_{\text{avg}}^{(2)}$ (e) and averaged to create $H_{\text{avg}}^{(3)}$ (f). Finally, $H_{n_4}$ is aligned with $H_{\text{avg}}^{(3)}$ (g) and averaged to generate the final interpolated track $H_{\text{avg}}^{(4)}$ depicted in (h). For visual clarity, panels (b)--(h) display of a single parameter, but the alignment is computed over the alignment parameter subset $\mathcal{A}$ and the resulting warping path is applied to all parameters simultaneously. \label{fig:overview}}
\end{figure*}

In this study, we propose a novel approach to binary stellar evolution track interpolation. Our method utilizes pre-simulated binary star populations with varied initial conditions (represented as grid $G$) to accurately predict evolutionary sequences for arbitrary initial conditions within the parameter space. We employ a KNN algorithm enhanced with sophisticated sequence alignment techniques to address the challenges of irregularly sampled evolutionary tracks. Figure \ref{fig:overview} presents the general workflow of our approach, which is detailed in the following subsections.
%%%%%%%%%%%%%%%%%%%%%%%%%%%%%%%%%%%%%%%
%%%%%%%%%%%%%%%%%%%%%%%%%%%%%%%%%%%%%%%

%%%%%%%%%%%%%%%%%%%%%%%%%%%%%%%%%%%%%%%

\subsection{K-Nearest Neighbor Interpolation}
For a given target initial value $i_* = (M_1^*, M_2^*, P^*)$, our approach first determines the $K$ nearest initial values within the simulation grid $G$. Traditional KNN-based methods employ distance metrics such as Euclidean distance:

\begin{equation} \label{eq:l2}
d_j = \sqrt{(M_1^* - M_1^j)^2 + (M_2^* - M_2^j)^2 + (P^* - P^j)^2}
\end{equation}
to quantify similarity and thus determine a set of neighbors $\mathcal{N}$. One way to determine the relative importance of each neighbor is to define weights according to $w_j = d_j^{-p} / {\sum \limits_{k \in \mathcal{N}} d_k^{-p}}$ where $p=\{1,2,3,\ldots\}$.

Alternatively, barycentric weights derived from Delaunay triangulation can also be effectively applied to identify neighbors and determine their relative weights in the interpolation process. To calculate barycentric weights \cite{berrut_barycentric_2004}, we construct a convex hull over the grid using Delaunay triangulation \cite{barber_quickhull_1996}. Each vertex in the triangulation corresponds to an initial condition $i_j$ from grid $G$. For any target initial condition $i_*$, we identify the tetrahedron containing this point and compute its barycentric coordinates with respect to the tetrahedron vertices. This process yields a set of neigbours $\mathcal{N}^* = \{n_1, n_2, \ldots, n_j, \ldots, n_K\}$ where $n_j$ are track indices with corresponding barycentric weights 
\begin{equation}
B^* = \{b_1, b_2, \ldots, b_j, \ldots, b_K\}, \ \ \  \sum_{j=1}^{K} b_j = 1,
\end{equation}
where $K=4$ for our 3D grid. Our experimental evaluation demonstrates substantial performance improvements when using barycentric weights $B$ compared to traditional Euclidean distance-based weighting schemes. In cases where the target initial value does not fall within any tetrahedron, no neighbors are found using barycentric coordinates. When this occurs, our algorithm identifies the nearest neighbor track using Equation \ref{eq:l2} and uses it directly as the prediction.

After identifying the most similar neighbors and determining their respective weights, we retrieve the corresponding evolutionary tracks $H_j$. The interpolation process then proceeds to track alignment before the final averaging operation.

%%%%%%%%%%%%%%%%%%%%%%%%%%%%%%%%%%%%%%%
%%%%%%%%%%%%%%%%%%%%%%%%%%%%%%%%%%%%%%%
\begin{figure*}
%\plotone{figs/dtw_v4.pdf}
\includegraphics[width = \textwidth]{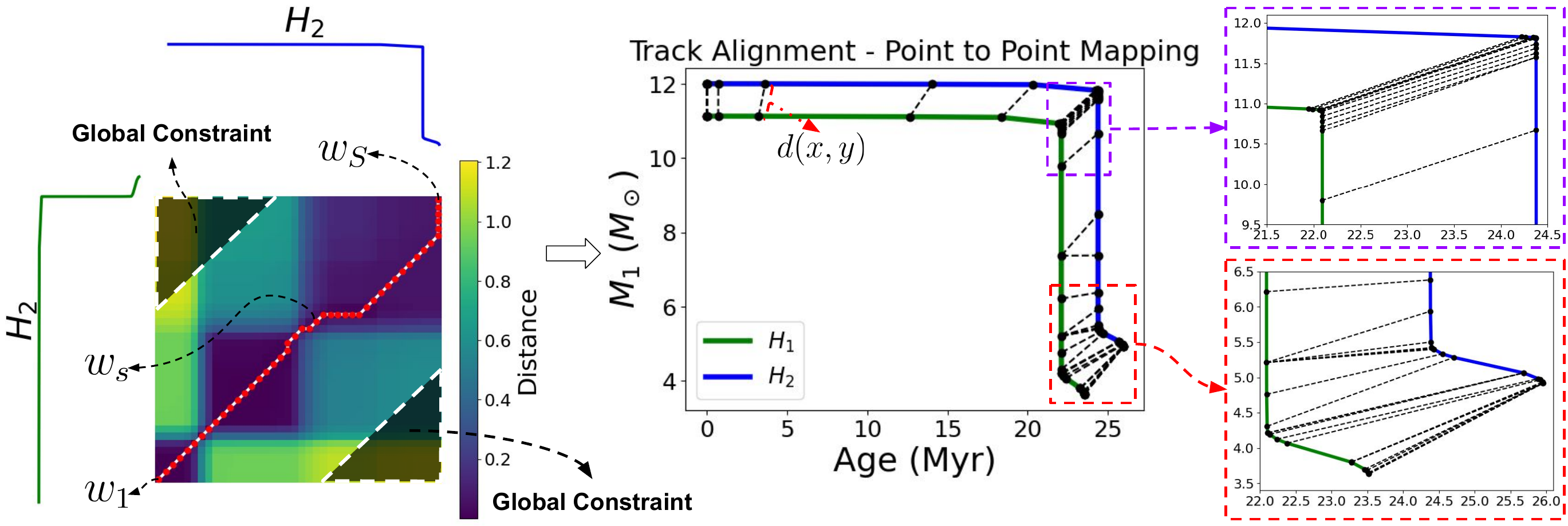}
\caption{Dynamic Time Warping alignment process. (Left) The pairwise distance matrix between each point of tracks $H_A$ and $H_B$, with color-coding indicating distance values (lighter colors represent higher values). DTW finds the optimal warping path (red points) that minimizes the cumulative distance function (Equation \ref{eq:cum_dist}). The shaded regions illustrate the global constraints that exclude the constrained region from search areas. The global constraint is used in Constrained DTW (CDTW) to prevent excessive warping. (Middle) The aligned tracks show corresponding points connected by dashed lines. (Right) The close-up panel demonstrates the precise alignment of signal features, particularly at edges where there is a significant drop in parameter values.
\label{fig:dtw}}
\end{figure*}
%%%%%%%%%%%%%%%%%%%%%%%%%%%%%%%%%%%%%%%

\subsection{Dynamic Time Warping (DTW)} \label{sec:dtw}
Dynamic Time Warping (DTW) is an algorithm for aligning two temporal sequences by finding correspondences between time points that minimize a cumulative distance measure. DTW finds point-to-point correspondences between tracks without requiring uniform temporal sampling or equal sequence lengths. Given two evolutionary tracks;
\begin{equation}
\begin{split}
H_A &= \{(a_1^A, \mathbf{v}_1^A), (a_2^A, \mathbf{v}_2^A), \ldots, (a_x^A, \mathbf{v}_x^A), \ldots, (a_X^A, \mathbf{v}_X^A)\} \\
H_B &= \{(a_1^B, \mathbf{v}_1^B), (a_2^2, \mathbf{v}_2^B), \ldots, (a_y^B, \mathbf{v}_y^B), \ldots, (a_Y^B, \mathbf{v}_Y^B)\}
\end{split}
\end{equation}
where $\mathbf{v}^A_x, \mathbf{v}^B_y \in \mathbb{R}^D$ are vectors containing all tracked physical parameters at each time point and the tracks may have different numbers of time points ($X \neq Y$). DTW constructs an optimal warping path
%with potentially different numbers of time points $(X \neq Y)$, DTW constructs an optimal warping path 
\begin{equation}
W_{A,B} = \{w_1, w_2, \ldots, w_s \ldots, w_S\}    
\end{equation}
where each element $w_s = (x, y)$ represents a correspondence between the $x$-th point in $H_A$ and the $y$-th point in $H_B$. The warping path must satisfy the following constraints:

\begin{enumerate}
\item \textbf{Boundary condition}: The warping path must connect the initial and final points of both sequences, that is $w_1=(1,1)$ and $w_S=(X,Y)$.

\item \textbf{Monotonicity}: For consecutive warping path elements $w_{s-1}=(x', y')$ and $w_s=(x,y)$, the conditions $x' \leq x$ and $y' \leq y$ must hold to preserve temporal order.

\item \textbf{Step size restriction}: To prevent excessive time warping, consecutive elements must satisfy $x - x' \leq 1$ and $y - y' \leq 1$. Since $x$ and $y$ represent indices of the track sequences, this constraint ensures that consecutive warping elements cannot skip more than one time step in either sequence.
\end{enumerate}

Given these constraints, DTW finds the optimal warping path $W$ by minimizing the cumulative distance function $\psi(x,y)$ through dynamic programming:

\begin{equation} \label{eq:cum_dist}
\psi(x,y) = d(x,y) + \min(\psi(x-1, y-1), \psi(x-1, y), \psi(x, y-1)),
\end{equation}
where $d(x,y)$ denotes the distance between points $(a^A_x, \mathbf{v}^A_x)$ and $(a^B_y, \mathbf{v}^B_y)$. The distance is computed using the alignment parameter subset $\mathcal{A}$, that is,
\begin{equation}
d(x,y) = \sqrt{(\tilde{a}^A_x - \tilde{a}^B_y)^2 + \sum_{\ell \in \mathcal{A}} (\tilde{v}^{A}_{x,\ell} - \tilde{v}^{B}_{y,\ell})^2}
\end{equation}
where $\tilde{v}^A_{x,\ell}$ and $\tilde{v}^B_{y,\ell}$ denote the $\ell$-th components of the normalized parameter vectors $\mathbf{v}^A_x$ and $\mathbf{v}^B_y$ corresponding to the parameter with index $\ell$, and the sum runs over the alignment parameter indices in $\mathcal{A}$. We employ Euclidean distance as the default metric unless otherwise specified. The resulting warping path $W$ is then applied to align all $D$ parameters simultaneously, not just those used for computing the alignment. Consequently, all physical parameters within a binary system maintain a coherent temporal correspondence. Figure~\ref{fig:dtw} illustrates an example distance matrix for calculating the optimal warping path. The alignment operation is formulated as
%where $d(x,y)$ denotes the distance between points $(a_x^A, v_x^A)$ and $(a_y^B, v_y^B)$. We employ Euclidean distance as the default metric unless otherwise specified. Figure \ref{fig:dtw} illustrates an example distance matrix for calculating the optimal warping path. The alignment operation is formulated as
\begin{equation}
\begin{split}
    \overline{H}_A &= align(H_A, W_{A,B}) = \{(a_t^A, \mathbf{v}_t^A) | t=1,\ldots S\}\\
    \overline{H}_B &= align(H_B, W_{B,A}) = \{(a_t^B, \mathbf{v}_t^B) | t=1,\ldots S\}
\end{split}
\end{equation}
where the $\mathrm{align}$ function takes a track $H_A$ and a  warping path $W_{A,B}$ as inputs and returns the aligned track  $\overline{H}_A$ to track $H_B$. For each element $w_s = (x, y)$  in $W_{A,B}$, the aligned track $\overline{H}_A$ contains the complete parameter vector $\mathbf{v}^A_x$ at position $s$, preserving all $D$ physical quantities under the same temporal reindexing. To align $H_B$ to $H_A$, we use the function $W_{B,A}$, obtained by swapping the index order so that each element $w_s = (x, y)$ becomes $w_s = (y, x)$. The resulting $\overline{H}_A$ and $\overline{H}_B$ have $S$ time points after alignment, and because all parameters share a single warping path, their mutual physical relationships are maintained at every time step.

%where the $align$ function applies the warping path to the complete parameter vector $\mathbf{v}$ at each time point. For each element $w_s = (x,y)$ in $W_{A,B}$, the aligned track $\bar{H}_A$ contains the full parameter vector $\mathbf{v}^A_x$ at position $s$. This maintains that all parameters are reindexed according to the same warping path, preserving their mutual physical relationships at each time step.
%where the $align$ function takes a track $H_A$ and a warping path $W_{A,B}$ as inputs and returns the aligned track $\overline{H}_A$ to track $H_B$. To align $H_B$ to $H_A$, we use the same function with $W_{B,A}$, which is obtained by swapping the index order in $W_{A,B}$. Each element $w_s = (x, y)$ in $W_{A,B}$ is converted to $w_s = (y, x)$ to obtain $W_{B,A}$. The resulting $\overline{H}_A$ and $\overline{H}_B$ have $S$ number of time points after alignment.

A critical consideration when applying vanilla DTW is the input preprocessing step. Each track contains both parameter values and corresponding age values for each point in a track. DTW compares points using Euclidean distance. Since age values are typically in millions of years while parameter values are in tens or hundreds, the distance calculation is dominated by the age component. This scale difference causes DTW to focus primarily on temporal proximity while ignoring parameter values, which is crucial for proper morphological alignment. We address this by normalizing both time coordinates and parameter values to the range $[0,1]$ using statistics from grid $G$. While this normalization substantially improves alignment quality, certain evolutionary phases may require different weighting of time versus parameter similarity. To address this, we implement several DTW variants:

\begin{itemize}
\item \textbf{Constrained DTW (CDTW)} \cite{cdtw}: Restricts the warping path to remain within a predefined band around the diagonal. Figure \ref{fig:dtw} shows the restricted regions defined by global constraints. Shaded areas are excluded from the warping path search even if they contain numerically better warping paths. This prevents excessive temporal shifts while aligning the signals.

\item \textbf{Derivative DTW (DDTW)} \cite{ddtw}: Rather than operating on raw or normalized sequences, DDTW pre-processes the data by computing signal derivatives prior to alignment.

\item \textbf{Shape DTW (SDTW)} \cite{shapedtw}: Provides a flexible pre-processing framework for signal alignment. It divides the signals into sub-sequences and extracts temporal features such as derivatives, slopes, or gradient histograms. These features are concatenated to form a feature vector. The vanilla DTW is applied to find correspondences on features. The resulting warping path is subsequently mapped back to the original signal space.

\end{itemize}

This alignment procedure establishes physically meaningful correspondences between evolutionary states across tracks with different temporal sampling. Once the warping path $W$ is calculated, we utilize it to determine which time points from each track should be averaged during the interpolation process. We utilize the \texttt{tslearn} library \citep{tslearn} in parts of our DTW implementation.

%%%%%%%%%%%%%%%%%%%%%%%%%%%%%%%%%%%%%%%

\subsection{Iterative Weighted Averaging}
The proposed method employs DTW to align neighboring tracks before the final interpolation step. DTW is specifically designed for aligning two signals simultaneously, which presents a fundamental constraint when working with multiple neighboring tracks. When we need to align multiple tracks to each other, we can only process them in pairs. Hence, a strategic approach to handle the complete set of neighbors is required.

To address this limitation, we implement an iterative alignment and averaging strategy. The process begins by sorting the neighboring tracks based on their barycentric weights, where higher weights indicate greater similarity in their initial values. The first two neighboring tracks are aligned using DTW, followed by computing their barycentric-weighted average. This averaged track serves as the reference for alignment with the subsequent neighboring track. The procedure continues iteratively until all neighboring tracks have been used. We can formulate this operation as:

\begin{equation}
\begin{split}
H_{avg}^{(1)} & = b_1 \cdot H_{n_1}\\
H_{avg}^{(k)} & = align(H_{avg}^{(k-1)}, W_{(H_{avg}^{(k-1)},H_{n_k})})\\ 
              &+ b_k \cdot align(H_{n_k}, W_{(H_{n_k}, H_{avg}^{(k-1)})}) \quad k=2,3,..K,
\end{split}
\end{equation}
where $H_{avg}^{(k)}$ represents the averaged track after incorporating the first $k$ neighbors, $b_k$ denotes the barycentric weight of the $k$-th neighbor. The final interpolated track $H_*$ is given by $H_{avg}^{(K)}$ after incorporating all K neighbors. Figure \ref{fig:overview} illustrates the neighbor searching process in the grid and the iterative weighted averaging steps.

It is important to note that the warping path at each iteration step is computed using only the alignment parameters $\mathcal{A}$, but the weighted averaging operation is applied to all $D$ parameters using this shared warping path. Specifically, at step $k$, the warping path $W(H^{(k-1)}_{\rm avg}, H_{n_k})$ is determined from the alignment subset, and this path governs the temporal correspondence for computing the weighted average across all physical quantities. This design stands in contrast to an alternative approach where each parameter is aligned independently, producing a separate warping path and consequently a different age grid for each physical quantity within the same binary system. Such per-parameter alignment can introduce physical anomalies: for instance, changes in stellar radius or effective temperature may appear to precede the onset of mass transfer, and coupled quantities such as luminosity $L$, radius $R$, and effective temperature $T_{\rm eff}$ may fall on incompatible temporal grids, leading to violations of physical relationships such as the Stefan-Boltzmann law. Our joint-alignment formulation avoids these artifacts by placing all parameters on a single consistent age grid, preserving causal ordering of evolutionary events and the physical relationships between coupled quantities. Section~\ref{sec:physical_consistency} provides a quantitative analysis of this consistency, examining the time evolution of coupled quantities such as luminosity, radius, and effective temperature, as well as Hertzsprung-Russell diagram comparisons and the Stefan-Boltzmann relation.

%%%%%%%%%%%%%%%%%%%%%%%%%%%%%%%%%%%%%%%

\subsection{Class Aware Interpolation}
Binary star evolution exhibits distinct physical regimes characterized by different mass transfer behaviors and other evolutionary outcomes. These regimes fundamentally influence the morphology of evolutionary tracks, particularly during critical phases. As discussed in Section \ref{sec:background}, each binary star system is classified according to these characteristics into four main categories: NMT, SMT, UMT, and IMT. We remove all tracks classified as IMT to focus our analysis on the three primary evolutionary categories, consistent with the evaluation framework established by \cite{Srivastava_2025}. Treating all binary systems uniformly in the interpolation process can lead to physically inconsistent predictions when averaging tracks from systems with fundamentally different evolutionary pathways.

Following the methodology for initial/final interpolation described in \cite{fragos_posydon_2023}, we classify evolutionary tracks according to their mass transfer behavior using a k-nearest neighbors approach. We determine the optimal value of k through Monte Carlo cross-validation and apply inverse squared distance weighting ($d^{-2}$) to each of the k neighbors.

Within grid $G$, each initial value $i_j$ has an associated classification label $c_j$. Our algorithm begins by dividing the grid into three subgrids $G_c$ where $c \in \{NMT, SMT, UMT\}$. Each subgrid $G_c$ contains exclusively binary stars belonging to class $c$. We construct separate Delaunay triangulations for each subgrid $G_c$, to search neighbors within the same class exclusively.

Section \ref{sec:experiment} provides a comprehensive analysis of the effects of using class-aware interpolation compared to interpolation without classification.

%%%%%%%%%%%%%%%%%%%%%%%%%%%%%%%%%%%%%%%%%%%%%%%%%%%%%%%%%%%%%%%%%%%%%%%%%%%
%%%%%%%%%%%%%%%%%%%%%%%%%%%%%%%%%%%%%%%%%%%%
\begin{deluxetable*}{c cc|ccc|c}
\label{table:grid_stat}
\digitalasset
\tablewidth{0pt}
\tablecaption{Sample Distribution Across Binary Evolution Grids}
\tablehead{
\colhead{Grid} & \colhead{\# Samples} & \colhead{\# Samples (filtered)} & \colhead{\# NMT}& \colhead{\# SMT} & \colhead{\# UMT} & \colhead{\# Test Samples}}
\startdata
HMS-HMS &60533&55736&15497&18712&21527&2833\\
CO-HeMS &39477&32868&23611&8095&1162&2296\\
CO-HMS &23684&18777&6025&8338&4414&2368\\
\enddata
\tablecomments{The table shows the number of tracks in each grid before and after applying quality filters. Tracks classified as Initial Mass Transfer (IMT) and those with insufficient temporal resolution (fewer than two time points) were excluded from all analyses. The remaining tracks are categorized by their evolutionary class: No Mass Transfer (NMT), Stable Mass Transfer (SMT), and Unstable Mass Transfer (UMT).}
\end{deluxetable*}
%%%%%%%%%%%%%%%%%%%%%%%%%%%%%%%%%%%%%%%%%%%%

\section{Experiments} \label{sec:experiment}

\subsection{Dataset}
Our experimental evaluation utilizes binary star populations simulated by the MESA framework \cite{fragos_posydon_2023}. We conduct separate analyses on three distinct grids (HMS-HMS, CO-HeMS, CO-HMS) representing different evolutionary regimes. During pre-processing, we implement a quality filter to ensure data integrity throughout our experiments. We excluded failed simulations and systems with initial mass transfer from the grid.

Table \ref{table:grid_stat} summarizes the sample counts before and after applying quality filtering criteria, across each evolutionary class following pre-processing and for testing. The grid structure follows a uniform sampling strategy in logarithmic space ($\log_{10}$) for the three initial parameters: $M_1$, $M_2$, and $P$. Our test samples are drawn randomly from this parameter space to simulate realistic inference conditions encountered during practical applications. %This dataset serves as the foundation for all subsequent experimental evaluations.

\subsection{Evaluation}
Assessing the quality of interpolated binary evolution tracks presents significant methodological challenges due to the adaptive temporal sampling inherent in binary evolution simulations. The MESA simulator strategically focuses computational resources on evolutionary phases characterized by rapid parameter changes, resulting in high temporal resolution during critical transitions and relatively sparse sampling during quiescent periods. This adaptive sampling strategy creates evaluation complications, as even minor temporal misalignments between predicted and ground-truth tracks can yield excessively large error values despite morphological similarity. From an astrophysical standpoint, minor timing offsets of ~1,000 years in critical events such as mass transfer onset are observationally insignificant and fall well below detection thresholds. The primary concern is preserving the overall track morphology and mass transfer outcomes rather than exact temporal alignment.

We employ two evaluation approaches to capture different aspects of interpolation accuracy. The first approach treats the predicted track as a continuous signal by connecting time points with linear interpolation. For each ground truth time point, the corresponding value in the predicted signal is calculated from this interpolated representation. These calculated values are then compared with the ground truth values to assess accuracy. This method provides insight into how well our interpolation captures parameter values at the exact ground truth time points.

The second evaluation metric treats both ground truth and predicted sequences as piecewise linear functions connecting adjacent time points. This approach requires numerical integration across the entire temporal domain to compute the error values that account for the complete evolutionary history rather than discrete point comparisons. This approach is referred as \textit{area error} since the integration can be approximated by calculating the area between the curves. This error measurement offers a more holistic assessment of interpolation quality by considering the cumulative deviation between predicted and actual evolutionary pathways throughout the binary system's lifetime.

%Both evaluation methodologies provide complementary perspectives on interpolation performance. %, with point-wise comparison highlighting accuracy at critical evolutionary milestones and integrated metrics capturing overall trajectory fidelity. 
The following sections present the mathematical formulations for these error calculation approaches. In our experimental evaluation, we compare our DTW-based method against two baseline approaches: a simple nearest neighbor method that uses the single closest track from the grid as the prediction, and the Change Point algorithm developed by \cite{Srivastava_2025}.

%%%%%%%%%%%%%%%%%%%%%%%%%%%%%%%%%%%%%%%%%%%%
\begin{figure*}[h]
%\plotone{violin_abs.eps}
\includegraphics[width = \textwidth]{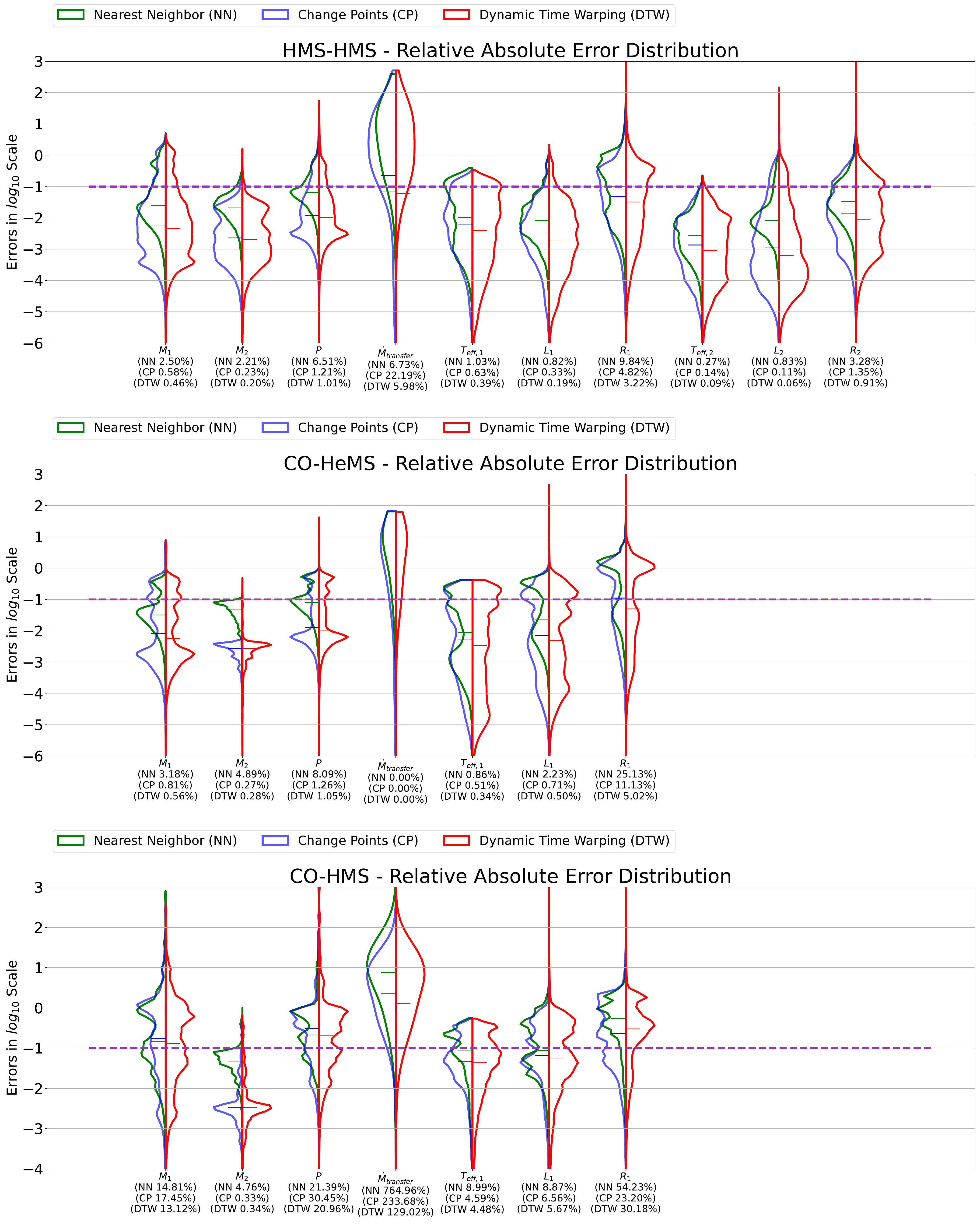}
\caption{Comparison of interpolation methods using Relative Absolute Error distributions. Violin plots show the distribution of RAE values for the Change Point, Nearest Neighbor, and our DTW-based iterative averaging methods across all binary evolution parameters. The y-axis is scaled logarithmically ($\log_{10}$) to visualize the full error range, while median values are presented as percentage Relative Absolute Error (shown in parentheses on the x-axis). The purple dashed line indicates the 10\% error threshold, above which predictions are considered unacceptable for astrophysical applications. Our DTW-based method consistently achieves lower median errors across most of the parameters for all 3 grids.
\label{fig:error_abs}}
\end{figure*}
%%%%%%%%%%%%%%%%%%%%%%%%%%%%%%%%%%%%%%%%%%%%

\subsection{Relative Absolute Error} \label{sec:rae}
The Relative Absolute Error (RAE) measures the absolute difference between prediction and ground truth values, normalized relative to the magnitude of the actual values. This metric effectively expresses the percentage deviation from the true value, providing a scale-invariant measure of prediction accuracy.

For any given initial condition $i_*$, our interpolation method $f$ generates a predicted track $H_* = f(i_*)$ where $H_* = \{(a_x^*, v_x^*) | x = 1, \ldots, X\}$. The corresponding ground truth sequence is denoted as $H_*^{GT} = \{(a_y^{GT}, v_y^{GT}) | y = 1, \ldots, Y\}$. Note that the number of time points in the prediction ($X$) and ground truth ($Y$) sequences may differ, and their temporal sampling points are generally not aligned.

%%%%%%%%%%%%%%%%%%%%%%%%%%%%%%%%%%%%%%%%%%%%
\begin{figure}
\begin{center}
%\plotone{sample_mtf.pdf}
\includegraphics[width = \linewidth]{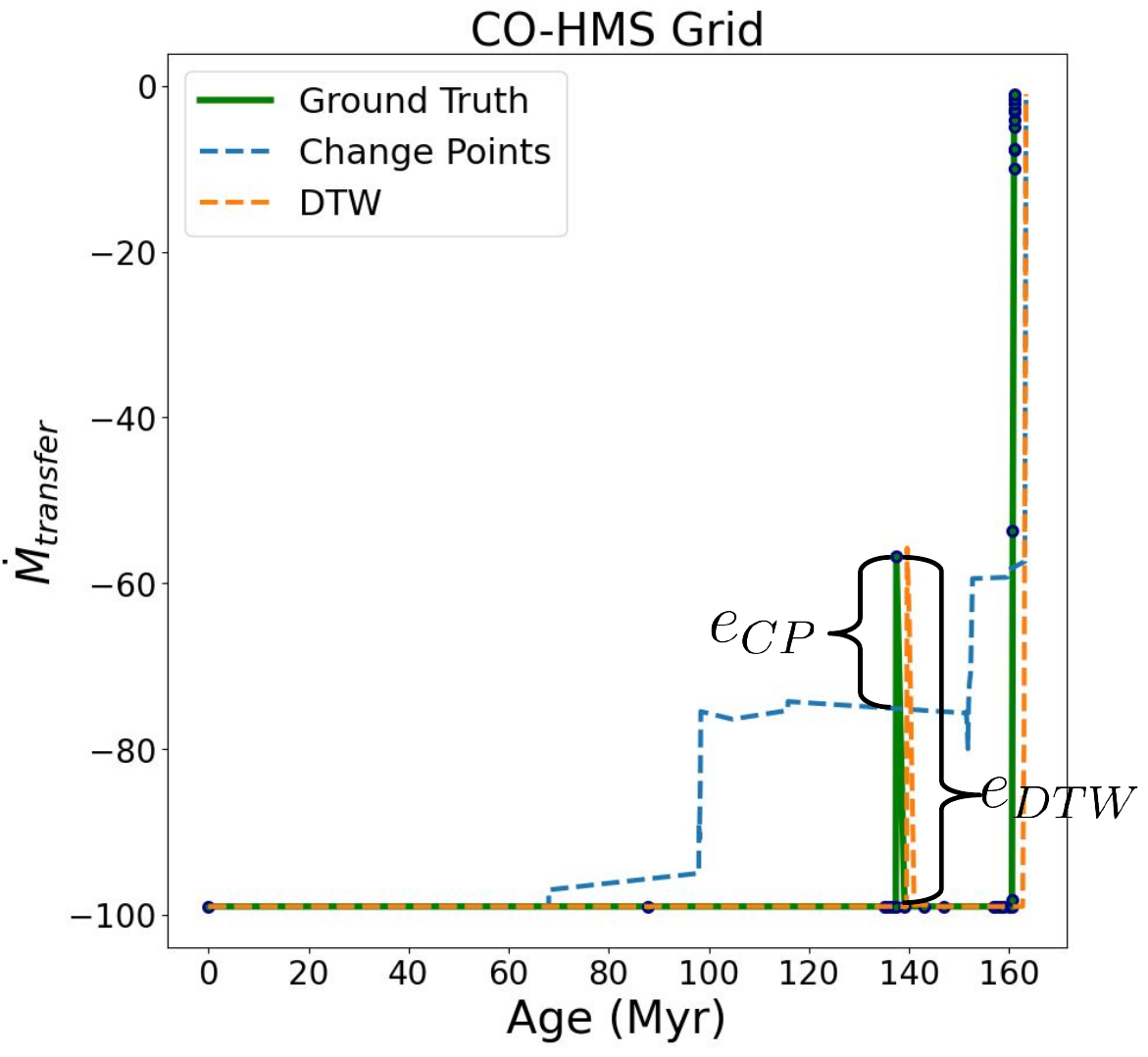}
\caption{Sample interpolation results for the $\dot{M}_{transfer}$ parameter from the CO-HMS grid with initial conditions $M_1$ = 4.31$M_{\odot}$, $M_2$ = 1.58$M_{\odot}$ and $P$ = 129.4 days. The ground truth (green solid line) is compared against predictions from the Change Point (blue dashed) and DTW (orange dashed) algorithms. Despite the DTW prediction being visually closer to the ground truth, it exhibits larger relative absolute error due to slight temporal shifts. At the first spike, the absolute error $e_{DTW}$ measures the distance from the spike peak to the interpolated value, while $e_{CP}$ is smaller despite the Change Point prediction missing the spike morphology. At the second spike, both methods are severely penalized due to temporal misalignment: the error is calculated from signal peaks (near 0) to minima (around $-100$), despite close visual agreement. The dense clustering of ground truth sampling points during rapid changes results in inflated RAE values, motivating the area-based error metric that better captures track fidelity.
\label{fig:sample_mtf}}
\end{center}
\end{figure}

%%%%%%%%%%%%%%%%%%%%%%%%%%%%%%%%%%%%%%%%%%%%

%%%%%%%%%%%%%%%%%%%%%%%%%%%%%%%%%%%%%%%%%%%%
\begin{figure*}
\begin{center}
%\plotone{figs/error_calculationv4.pdf}
\includegraphics[width = \textwidth]{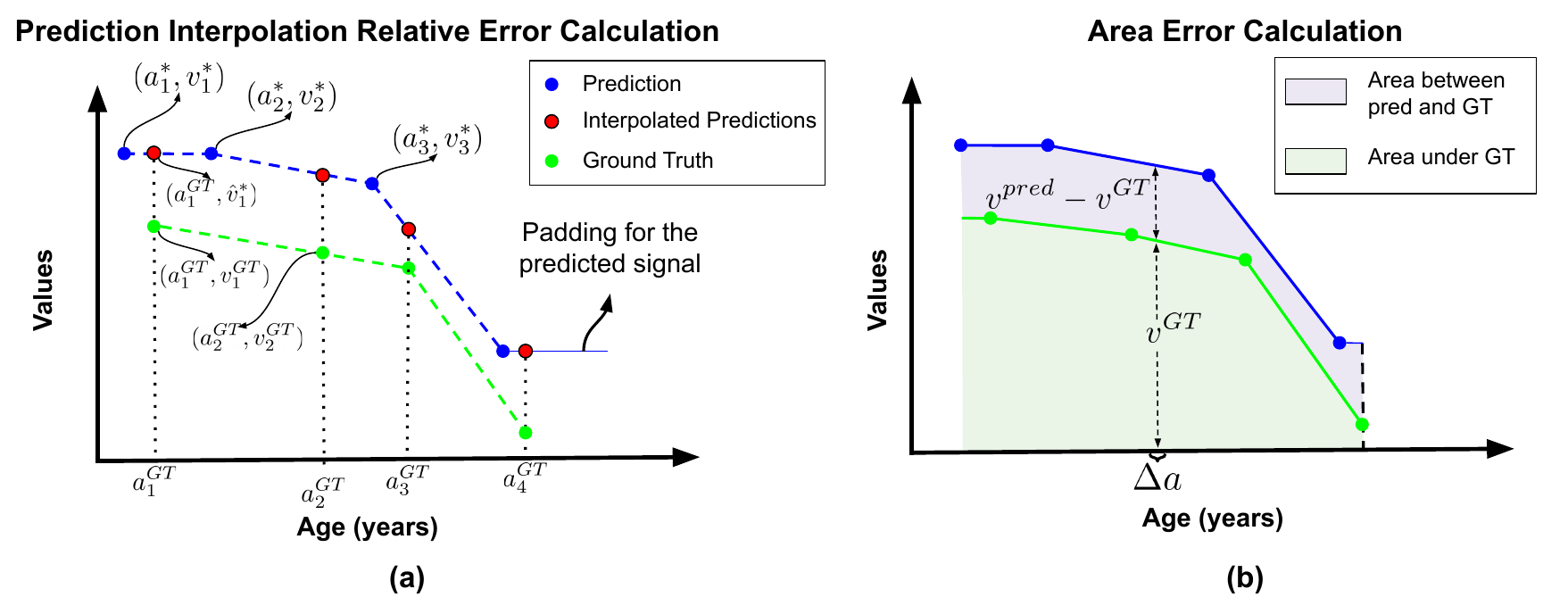}
\caption{Error calculation methodologies for evaluating interpolation accuracy. (a) Relative Absolute Error (RAE) computation: RAE evaluates deviations at specific ground truth time points. Since interpolated tracks are irregularly sampled and may not contain values at exact ground truth timestamps, we create a continuous representation using linear interpolation between adjacent points. The interpolated prediction is then sampled at ground truth time points to calculate relative deviations. Padding is applied at boundaries when predictions are shorter than ground truth tracks. (b) Area Error calculation: Both predicted time points (blue dots) and ground truth time points (green dots) are linearly interpolated to treat them as continuous signals. Area error quantifies prediction deviations across the entire temporal domain by computing the relative absolute distance between tracks at infinitesimally small time steps $\Delta a$. The calculation is approximated by measuring the area between the predicted and ground truth tracks (blue shaded region) and normalizing by the area under the ground truth curve (green shaded region).
\label{fig:error_calc}}
\end{center}
\end{figure*}

%%%%%%%%%%%%%%%%%%%%%%%%%%%%%%%%%%%%%%%%%%%%

Since the ground truth sequence provides reference time points $a_y^{GT}$ for $y = 1, \ldots, Y$, we evaluate the prediction accuracy exclusively at these temporal coordinates. To obtain prediction values at ground truth time points, we construct a continuous representation of $H_*$ through linear interpolation. Sampling this continuous function at the ground truth time points yields:

\begin{equation}
\hat{H}_* = \{(a_y^{GT}, \hat{v}_y^*) | y = 1, \ldots, Y\}
\end{equation}
where $\hat{v}_y^*$ represents the interpolated parameter value from the predicted track $H_*$ at ground truth time point $a_y^{GT}$. Figure \ref{fig:error_calc} illustrates this interpolation process for error computation. RAE is calculated at each ground truth time point using the formula:

%\begin{equation}
%\mathbb{E}_\text{RAE}(a_y) = \frac{|\hat{v}_y^* - v_y^{GT}|}{|v_y^{GT}|}.
%\end{equation}
\begin{equation}
\mathbb{E}_\text{RAE}(a_y) = \left| \frac{\hat{v}_y^* - v_y^{GT}}{v_y^{GT}} \right|.
\end{equation}

Rather than averaging the error for each track individually, we analyze the distribution of $\mathbb{E}_\text{RAE}$ values across all time points from all tracks in our evaluation set. %This approach provides a comprehensive view of prediction accuracy across the entire parameter space and evolutionary timeline. 

Figure \ref{fig:error_abs} presents violin plots of the distribution of errors among different methods. The RAE values are displayed in logarithmic ($\log_{10}$) scale to clearly depict the distribution characteristics. Median values are shown for each distribution on the plot along with their percentage Relative Absolute Error. The percentage Relative Absolute Error values for each algorithm are evaluated against a 10\% threshold that astronomers consider critical for acceptable accuracy. The error distributions reveal that our proposed method consistently outperforms the Change Point and Nearest Neighbor algorithms across all parameters for HMS-HMS and CO-HeMS grids. We observe particularly dramatic improvements for the logarithmic mass transfer rate $\dot{M}_{transfer}$ parameter due to its characteristically sharp, spike-like variations. However, the CO-HMS grid presents different behavior. While our DTW-based method achieves the lowest median errors for $M_1$ and most other parameters, the Change Point approach yields better results for $R_1$. Additionally, all three methods exhibit substantial errors for $\dot{M}_{\mathrm{transfer}}$, which can be attributed to the distinctive characteristics of mass transfer rates in CO-HMS systems. Nevertheless, our method reduces the $\dot{M}_{\mathrm{transfer}}$ median error by nearly half compared to the Change Point algorithm.%However, the CO-HMS grid presents different behavior. The NN method performs better for the $M_1$ parameter, while the Change Point approach yields better results for $R_1$. Additionally, all three methods exhibit substantial errors for $\dot{M}_{transfer}$, which can be attributed to the distinctive characteristics of mass transfer rates in CO-HMS systems. 
Figure \ref{fig:sample_mtf} demonstrates that our predictions are acceptable and outperform the Change Point. RAE metric fails to fully capture the quality of the interpolated tracks. In the following section, we introduce an alternative metric that better reflects prediction quality for binary evolution tracks.

\subsection{Area Error} \label{sec:area_error}
While RAE provides valuable insight into prediction accuracy at specific time points, it has a significant limitation. It only evaluates error at ground truth timestamps without considering the predicted curve behavior between these points. This limitation means that two interpolated tracks with completely different morphologies could achieve identical RAE scores if they merely pass through the same values at evaluation timestamps.

%%%%%%%%%%%%%%%%%%%%%%%%%%%%%%%%%%%%%%%%%%%%
\begin{figure*}
%\plotone{violin_area.eps}
\includegraphics[width = \textwidth]{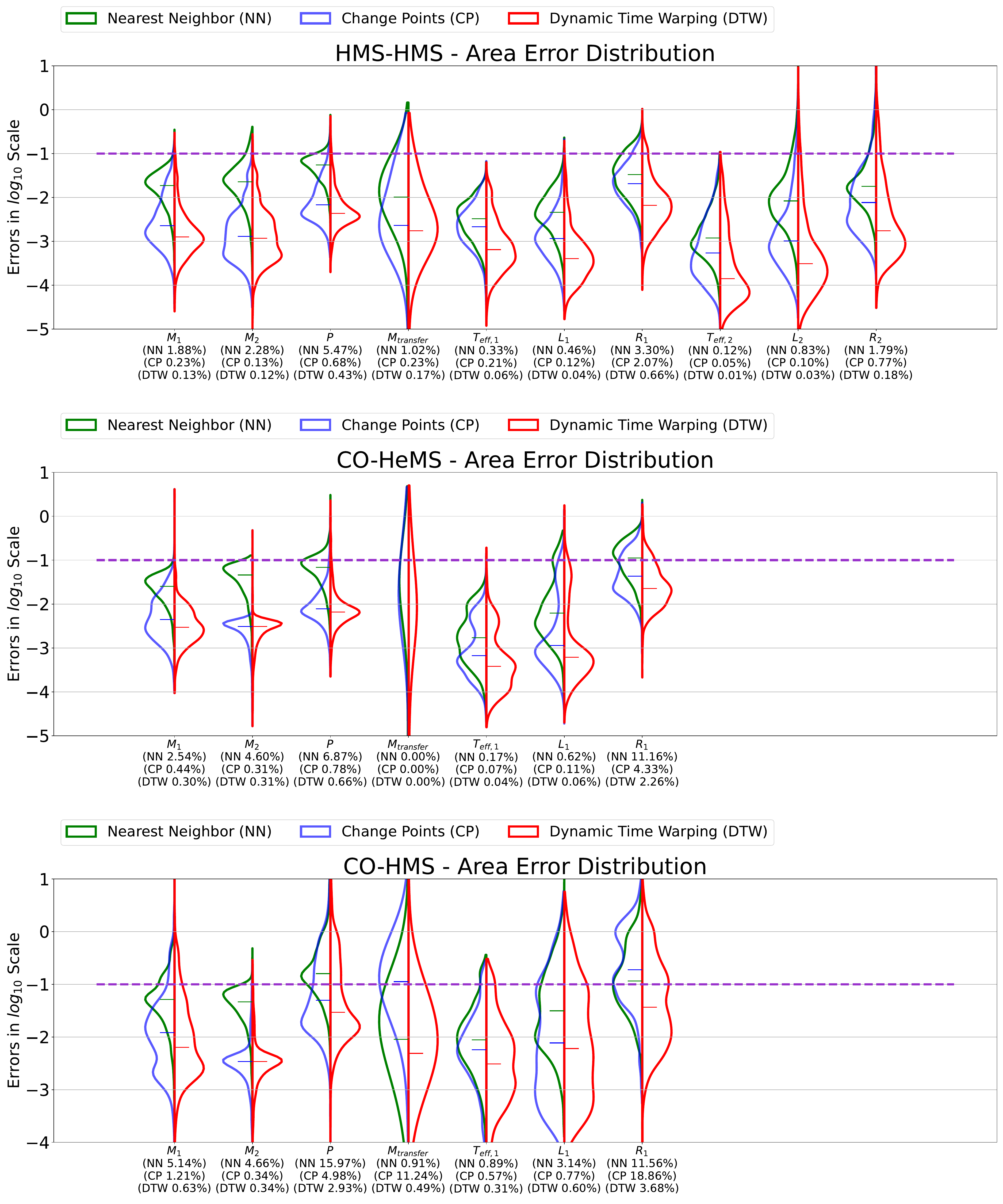}
\caption{Comparison of interpolation methods using Area Error distributions. Violin plots display the area error distributions for the Nearest Neighbor approach, Change Point algorithm, and our DTW-based iterative averaging method across all binary evolution parameters and three grids (HMS-HMS, CO-HeMS, and CO-HMS). The y-axis is scaled logarithmically, with median values marked within each distribution. Our DTW-based method demonstrates superior performance across all three grids, achieving consistently lower median area errors. \label{fig:error_area}}
\end{figure*}
%%%%%%%%%%%%%%%%%%%%%%%%%%%%%%%%%%%%%%%%%%%%

To address this limitation and obtain a more comprehensive understanding of prediction quality, we treat both predicted and ground truth tracks as continuous signals. Both predicted time points and ground truth time points are linearly interpolated to create continuous representations. The area error method calculates the area between the two curves. By normalizing this area against the area under the ground truth curve, we obtain a relative measure of prediction error that accounts for the complete evolutionary trajectory rather than just discrete points.
The mathematical formulation for area error is:

\begin{equation}
\mathbb{E}_\text{area} = \int_{}^{} \mathbb{E}_\text{area}(a) \,da \approx \frac{\int |H_* - H_*^{GT}|da}{\int |H_*^{GT}|da}
\end{equation}
where the integration is performed over the temporal range from the first to the last ground time point. Figure \ref{fig:error_calc} illustrates the computational process for calculating area error. Unlike $\mathbb{E}_\text{RAE}$, which produces a distribution of errors across all time points, $\mathbb{E}_\text{area}$ generates a single error value for each track. %, providing a track-specific assessment of overall prediction quality.

Figure \ref{fig:error_area} presents violin plots displaying the area error distribution across different track parameters for various grids, with values presented in logarithmic scale. The results demonstrate that the DTW-based approach achieves superior performance across all three grids (HMS-HMS, CO-HeMS, and CO-HMS), exhibiting consistently lower median area errors compared to the Change Point and Nearest Neighbor algorithms. In contrast to the RAE metric, the Area Error metric successfully captures the quality improvements of our approach for the CO-HMS grid.

%\pagebreak
\subsection{Visual Evaluation}
\begin{figure*}
%\plotone{sample1v6.pdf}
\centering
\includegraphics[width = 0.8\textwidth]{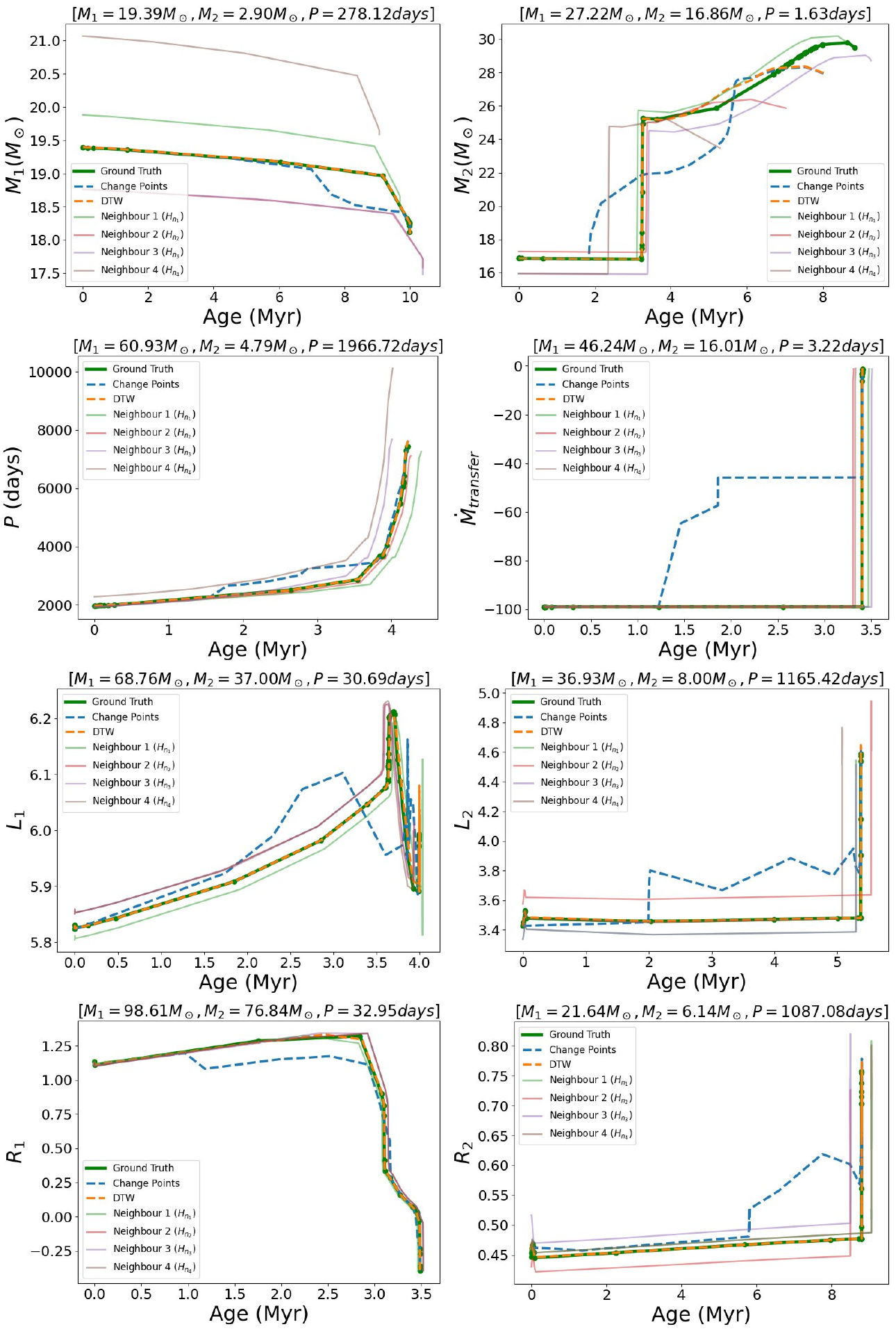}
\caption{Interpolation results comparing methods across different binary parameters. Initial conditions ($M_1, M_2, P$) for each sample are displayed in the panel titles. Each plot displays the neighboring tracks $H_{n_i}$ (used for interpolation), predictions from the Change Point algorithm (\cite{Srivastava_2025}), predictions from our DTW-based method, and the corresponding ground truth evolutionary track. The Change Point algorithm shows significant deviations from the ground truth, producing morphologically inconsistent predictions despite the similar shapes of neighboring tracks. In contrast, our DTW-based alignment approach achieves near-perfect overlap with the ground truth. The visual results demonstrate that proper temporal alignment is crucial for accurate interpolation.
\label{fig:sample1}}
\end{figure*}

While quantitative error metrics provide objective measures for comparing interpolation methods, visual examination offers invaluable insight into how our proposed approach improves upon existing techniques. Figure \ref{fig:sample1} presents representative examples comparing our DTW-based alignment method with the Change Point approach.% These examples include neighboring tracks with their respective barycentric weights, where higher weights indicate closer proximity in the parameter space, resulting in greater influence during the averaging operation. %The visual comparisons display clear morphological similarities between neighboring tracks, enabling intuitive assessment of expected predictions. 
The examples clearly show the shape patterns shared among neighboring tracks, indicating what the predicted results should look like. The Change Point algorithm produces deformations that fail to capture the ground truth characteristics, while the DTW-based method achieves near-perfect overlap with reference tracks. This discrepancy stems from misalignment of critical evolutionary features across neighboring tracks. %When change points align correctly, the Change Point algorithm performs acceptably. However, deformations arise in regions of misalignment where the algorithm fails to identify corresponding evolutionary phases across neighboring tracks.

Our iterative averaging approach addresses these alignment challenges by finding proper correspondence between time points. This methodology preserves morphological features and demonstrates superior performance for the tracks exhibiting sudden parameter changes. The improvement is especially pronounced for logarithmic mass transfer rate interpolation $\dot{M}_{transfer}$, as shown in Figure \ref{fig:sample1}.

\begin{figure*}
%\plotone{figs/sample2.pdf}
\includegraphics[width = \textwidth]{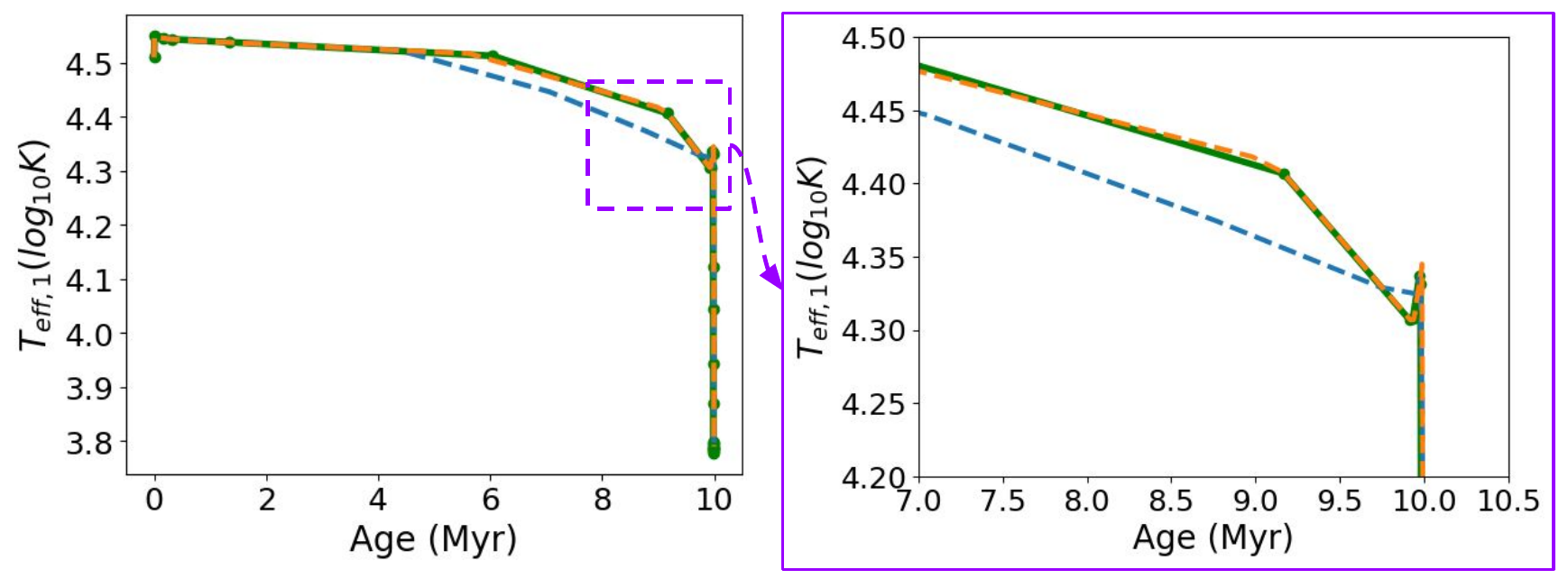}
\caption{Example interpolation results for $\log_{10}(T_{\text{eff},1})$ parameter with initial conditions $M_1$ = 19.39$M_{\odot}$, $M_2$ = 2.90$M_{\odot}$ and $P$ = 278.12 days. The close-up section highlights DTW-based approach capturing fine-scale variations in the signal, while the Change Point algorithm fails to preserve these critical details.
\label{fig:sample2}}
\end{figure*}

Figure \ref{fig:sample2} provides a closer look at interpolated signals, demonstrating the proposed method's ability to capture fine-grained variations. In contrast, the Change Point approach produces noticeable deviations in misaligned regions and fails to preserve subtle temporal features.

%%%%%%%%%%%%%%%%%%%%%%%%%%%%%%%%%%%%%%%%%%%%

%%%%%%%%%%%%%%%%%%%%%%%%%%%%%%%%%%%%%%%%%%%%
%%%%%%%%%%%%%%%%%%%%%%%%%%%%%%%%%%%%%%%%%%%%

%%%%%%%%%%%%%%%%%%%%%%%%%%%%%%%%%%%%%%%%%%%%%%%%%%%%%%

\begin{deluxetable*}{lccccccccccc}[t]
\label{table:ablation}
\digitalasset
\tablewidth{0pt}
\tablecaption{Ablation Study Results on HMS-HMS Grid}
\tablehead{
\colhead{Method} & \colhead{Class-Aware} & \colhead{$M_1$} & \colhead{$M_2$} & \colhead{$P$}& \colhead{$\dot{M}_{transfer}$} & \colhead{$T_{eff,1}$} & \colhead{$T_{eff,2}$}& \colhead{$L_1$} & \colhead{$L_2$} & \colhead{$R_1$} & \colhead{$R_2$}
}
\startdata
Change Points (4) & \checkmark & 1.8122 & 0.3381 & 3.2128 & 33.675 & 1.4933 & 0.3972 & 1.1133 & 0.4695 & 9.716 & 4.036  \\
Change Points (8) & \checkmark  & 0.9217 & 0.2486 & 1.8003 & \textcolor{blue}{22.1917}  & 0.9608 & 0.2417 & 0.5851  & 0.2386 & 8.0558 & 2.0624 \\
Change Points (16) & \checkmark  & 0.596 & \textcolor{blue}{0.2277} & 1.2938 & 23.2439 & 0.7345 & 0.1712 & 0.4104 & 0.1846 & 5.9181 & 1.5825 \\
Change Points (32) & \checkmark  & \textcolor{blue}{0.5846} & 0.2327 & \textcolor{blue}{1.2103} & 36.3708 & \textcolor{blue}{0.6268}  & 0.1464 & \textcolor{blue}{0.3267} & 0.1192 & \textcolor{blue}{4.8179} & 1.4481 \\
Change Points (64) & \checkmark  & 0.6882 & 0.2332 & 1.3281 & 39.5013  & 0.6803 & \textcolor{blue}{0.1358} & 0.3285 & \textcolor{blue}{0.1103} & 5.2436 & \textcolor{blue}{1.3505} \\
Change Points (128) & \checkmark  & 0.7585  & 0.247  & 1.4474 & 38.4068 & 0.7273 & 0.1462 & 0.3645 & 0.134 & 6.0697 & 1.492 \\
\hline
DTW (0-1 norm) & \checkmark  & \textcolor{red}{0.4588} & 0.2016 &  1.0146 & \textcolor{red}{5.9789} & \textcolor{red}{0.3935} & \textcolor{red}{0.0898} & \textcolor{red}{0.1948} & 0.0616 & \textcolor{red}{3.2208} & \textcolor{red}{0.9075} \\
DTW (max norm) & \checkmark  & 0.4742 & 0.2058 &  1.0458 & 6.4124 & 0.4206 & 0.0935 & 0.2116 & 0.0646 & 3.3949 & 0.9241 \\
CDTW (radius:3) & \checkmark & 0.4642 & 0.2015 &  1.0214 & 6.5293 & 0.3998 & 0.0909 & 0.1965 & 0.0625 & 3.264  & 0.915  \\
DDTW & \checkmark  & 0.4856 & \textcolor{red}{0.1989} &  1.0354 & 7.6014 & 0.407  & 0.0926 & 0.2068 & \textcolor{red}{0.0615} & 3.3858 & 0.9124 \\
SDTW & \checkmark  & 0.4721 & 0.2037 &  1.0415 & 6.945  & 0.4254 & 0.0931 & 0.2173 & 0.0662 & 3.4092 & 0.9374 \\
\hline
DTW (0-1 norm) &  & 0.5059 & 0.2062 & \textcolor{red}{1.012}  & 8.3602 & 0.4667 & 0.1018 & 0.2254 & 0.0678 & 3.6976 & 1.0406 \\
DTW (max norm) &  & 0.5019 & 0.2099 & 1.014  & 8.2953 & 0.4583 & 0.1042 & 0.2312 & 0.0735 & 3.6398 & 1.0527 \\
CDTW (radius:3) &  & 0.5147 & 0.2067 & 1.0212 & 9.0583 & 0.4742 & 0.1028 & 0.2278 & 0.0688 & 3.7314 & 1.0479 \\
DDTW &  & 0.5403 & 0.2025 & 1.0702 & 9.5391 & 0.4634 & 0.1    & 0.2324 & 0.0702 & 3.7186 & 1.0132 \\
SDTW &  & 0.5093 & 0.2071 & 1.0435 & 8.8804 & 0.4845 & 0.1053 & 0.2405 & 0.0731 & 3.8062 & 1.0675 \\
\enddata
\tablecomments{Performance comparison across different hyperparameters and DTW variants. \textcolor{blue}{Blue} indicates the best performing configuration for the Change Point algorithm, while \textcolor{red}{red} shows the best results for our DTW-based method. Our approach consistently outperforms the Change Point algorithm across all binary evolution parameters. Class-aware interpolation generally achieves better accuracy than methods that ignore it. The optimal hyperparameters and DTW variants vary across different parameters due to the distinct signal characteristics of each physical quantity. (0-1 norm) indicates input normalization where track values and ages are normalized between 0 and 1. (max norm) indicates normalization by dividing values by the maximum value. (radius) is the hyperparameter for CDTW, indicating the alignment constraint radius as explained in the methods section.}
\end{deluxetable*}
%%%%%%%%%%%%%%%%%%%%%%%%%%%%%%%%%%%%%%%%%%%%%%%%%%%%%%

%%%%%%%%%%%%%%%%%%%%%%%%%%%
\subsection{Ablation Study}
\subsubsection{Dynamic Time Warping Variants}
This section investigates performance variations resulting from different hyperparameter configurations for both the baseline Change Point method and the proposed DTW-based iterative weighted averaging approach. For the Change Point algorithm, we focus on its most critical hyperparameter: the number of change points, which determines the standardized track length after selecting the most representative points from each signal. %For the Change Point algorithm, we implemented class-aware subgrids exactly as prescribed by \cite{philipp}.

For our method, we examine the performance of DTW variants, including vanilla DTW, Constrained DTW (CDTW), Derivative DTW (DDTW), and Shape DTW (SDTW). Additionally, we evaluate the impact of class-aware interpolation by comparing performance with and without evolutionary class constraints.

Given the impracticality of presenting comprehensive visual comparisons for all configurations, we report median values from error distributions as our primary performance metric. Lower median values indicate superior performance across the evaluation set. Table \ref{table:ablation} summarizes the ablation study results, demonstrating that our proposed method consistently outperforms the Change Point algorithm across all parameters. This highlights the vital importance of proper track alignment for accurate interpolation.

When employing the class-aware approach, any DTW variant outperforms the Change Point algorithm. Incorporating class information significantly improves performance in DTW-based methods because tracks within the same evolutionary class generally exhibit similar morphological characteristics. This advantage is most pronounced near class boundaries. For samples located away from class borders, the neighbor selection remains largely unchanged whether class information is used or not. However, for samples near class boundaries, utilizing class labels results in a different set of neighbors, leading to more consistent interpolations and improved accuracy.

Another factor that affects DTW-based methods is input normalization. DTW calculates distances between time points across two tracks, where each time point contains both a parameter value and an age indicating the temporal measurement. Compared to parameter values, age has substantially larger magnitudes—while parameter values may be in the hundreds, ages can reach millions of years. This scale difference significantly impacts alignment performance. Without normalization, age becomes the dominant term in distance calculations, resulting in improper alignment that prioritizes temporal proximity over parameter similarity.

To address this issue, we normalize tracks before calculating the warping path. We tested two normalization approaches: (i) 0-1 normalization, where parameters and ages are separately normalized between 0 and 1, and (ii) max normalization, where values are divided by the maximum value from the training set. Results in Table \ref{table:ablation} show that 0-1 normalization generally yields better performance across most parameters compared to max normalization, particularly when combined with class-aware interpolation.
%Results in Table \ref{table:ablation} show that different normalization methods yield varying performance across parameters. This variation occurs because each parameter has different value ranges, and the relative importance between age and parameter value differs based on the morphological characteristics of each track.

%%%%%%%%%%%%%%%%%%%%%%%%%

\subsubsection{Iterative Weighted Averaging}

The proposed iterative weighted averaging approach sorts the $K$ neighbors based on their barycentric weights $b_j$ as illustrated in Figure \ref{fig:overview}. This approach is necessary because DTW can only align two tracks simultaneously, requiring us to determine the order in which tracks are aligned and averaged. Intuitively, we attempted to start with the most similar neighbors ($b_j$ are sorted in descending order) and continue until all neighbor tracks are incorporated. Since alignment proceeds with the averaged signal and the next neighbor, the order of track alignment can impact performance.

In this ablation study, we investigated different sorting strategies for the iterative weighted averaging process. Table \ref{table:ablation_neig_order} compares the performance of ordering neighbors based on $b_j$ in descending, ascending, and random order across different DTW variants. The results show that neighbor ordering does not significantly affect performance. While the best results for each DTW variant are highlighted in color and show no substantial differences, the optimal ordering varies across DTW variants: descending order performs best for max normalization, while ascending order shows 
marginal advantages for 0-1 normalization. Since we compute a weighted average at each step, the most similar signals can have a greater impact on the final outcome. The minimal sensitivity to ordering suggests that the iterative averaging procedure is robust to the sequence in which neighbors are incorporated.
%The results show that neighbor ordering does not significantly affect performance. While the best results for each DTW variant are highlighted in color and show no substantial differences, descending order yields slightly better results in most cases. This outcome aligns with our intuition that descending order, which starts the alignment from the most similar signals, produces better results. Since we compute a weighted average at each step, the most similar signals can have a greater impact on the final outcome. Placing the least similar signals last allows the algorithm to filter out potential outliers that might otherwise distort the interpolation.

%%%%%%%%%%%%%%%%%%%%%%%%%%%%%%%%%%%%%%%%%%%%%%%%%%
\begin{deluxetable*}{lc ccc c cc cc cc}[t]
\label{table:ablation_neig_order}
\digitalasset
\tablewidth{0pt}
\tablecaption{Effect of Neighbor Ordering on Iterative Weighted Averaging Performance on HMS-HMS Grid}
\tablehead{
\colhead{Method} & \colhead{Neighbor Order} & \colhead{$M_1$} & \colhead{$M_2$} & \colhead{$P$}& \colhead{$\dot{M}_{transfer}$} & \colhead{$T_{eff,1}$} & \colhead{$T_{eff,2}$}& \colhead{$L_1$} & \colhead{$L_2$} & \colhead{$R_1$} & \colhead{$R_2$}}
\startdata
DTW (0-1 norm)  & Descending & 0.4588 & 0.2016 & \textcolor{red}{1.0146} & 5.9789 & 0.3935 & \textcolor{red}{0.0898} & \textcolor{red}{0.1948} & 0.0616 & \textcolor{red}{3.2208} & 0.9075 \\
DTW (0-1 norm)  & Ascending  & \textcolor{red}{0.4554} & \textcolor{red}{0.2005} & 1.0168 & \textcolor{red}{5.7059} & \textcolor{red}{0.3913} & 0.09   & 0.1958 & \textcolor{red}{0.0613} & 3.2217 & \textcolor{red}{0.9033} \\
DTW (0-1 norm)  & Random     & 0.4615 & 0.2017 & 1.0177 & 5.7902 & 0.3959 & 0.0905 & 0.1958 & 0.0613 & 3.2371 & 0.912  \\
\hline
DTW (max norm) & Descending & \textcolor{red}{0.4742} & \textcolor{red}{0.2058} & \textcolor{red}{1.0458} & \textcolor{red}{6.4124} & \textcolor{red}{0.4206} & \textcolor{red}{0.0935} & \textcolor{red}{0.2116} & \textcolor{red}{0.0646} & \textcolor{red}{3.3949} & \textcolor{red}{0.9241} \\
DTW (max norm) & Ascending  & 0.4906 & 0.2107 & 1.0516 & 6.8531 & 0.437  & 0.0971 & 0.2261 & 0.0662 & 3.4889 & 0.9666 \\
DTW (max norm) & Random     & 0.4919 & 0.2091 & 1.0521 & 6.4917 & 0.4273 & 0.0981 & 0.2175 & 0.0651 & 3.4644 & 0.9659 \\
\hline
SDTW & Descending & 0.4721 & \textcolor{red}{0.2037} & 1.0415 & 6.945  & 0.4254 & \textcolor{red}{0.0931} & 0.2173 & \textcolor{red}{0.0662} & \textcolor{red}{3.4092} & \textcolor{red}{0.9374} \\
SDTW & Ascending  & 0.4719 & 0.2042 & \textcolor{red}{1.0299} & \textcolor{red}{6.8595} & \textcolor{red}{0.4251} & 0.094  & \textcolor{red}{0.2158} & 0.0669 & 3.4262 & 0.9541 \\
SDTW & Random     & \textcolor{red}{0.4707} & 0.2052 & 1.0386 & 6.97   & 0.4334 & 0.0948 & 0.217  & 0.0663 & 3.4608 & 0.9474 \\
\enddata
\tablecomments{Comparison of median relative absolute error values across different neighbor weight ($b_j$) ordering strategies (descending, ascending, and random) for DTW variants. Results are shown in $\log_{10}$ scale with the best performance for each DTW variant highlighted in \textcolor{red}{red}. Descending order sorts neighbors by decreasing barycentric weights (most similar first), ascending order sorts by increasing weights (least similar first), and random order uses an arbitrary sequence. The results demonstrate that neighbor ordering has minimal impact on overall performance. For some DTW variants, descending order shows slight advantages, while for others ascending order performs comparably or marginally better.%The results demonstrate that neighbor ordering has minimal impact on overall performance, though descending order shows slight advantages in most cases.
}
\end{deluxetable*}
%%%%%%%%%%%%%%%%%%%%%%%%%%%%%%%%%%%%%%%%%%%%%%%%%%

%%%%%%%%%%%%%%%%%%%%%%%%%

\subsubsection{Aligning Change Points}

In this section, we investigate the alignment capability of the Change Point algorithm. Each binary star evolutionary track in our dataset contains a different number of time points, requiring proper alignment before averaging similar neighboring tracks. The Change Point algorithm selects a predefined number of points based on importance criteria, assuming that these selected points preserve track morphology and provide adequate alignment. We argue that Change Points can miss critical time points and do not guarantee proper track alignment, as demonstrated by our visual results.

We further investigate the performance gap between our method and the Change Point algorithm to demonstrate that the Change Point approach can benefit from our DTW-based alignment. If Change Points provided better alignment than our method, applying DTW should not improve performance. However, our findings in Table \ref{table:ablation_cp_dtw} show that applying DTW to change points yields performance improvements across all experiments. Note that since the Change Point algorithm selects a different set of representative points for each parameter independently, the DTW alignment in this experiment is applied on a per-parameter basis.

Table \ref{table:ablation_cp_dtw} compares Change Point results with and without alignment. When examining each different number of change points individually, DTW variants consistently improve results. The best overall results for all parameters are also obtained using DTW alignment. Comparing the best results from Table \ref{table:ablation} and Table \ref{table:ablation_cp_dtw}, our proposed method outperforms the Change Point approach.

This demonstrates that Change Point selection discards important time points from the tracks. Since our method operates on original track values, it avoids information loss and achieves better performance. One might argue that Change Point results could improve by using more change points, but Table \ref{table:ablation_cp_dtw} shows that increasing the number of change points without alignment does not yield better performance. In fact, performance decreases because using more change points creates additional alignment issues.

%This study demonstrates that proper alignment is essential for effective track interpolation and highlights the importance of our DTW-based alignment method.

\begin{deluxetable*}{ll ccc c cc cc cc}[h!]
\label{table:ablation_cp_dtw}
\digitalasset
\tablewidth{0pt}
\tablecaption{Performance Improvement of Change Point Algorithm When Enhanced with DTW alignment on HMS-HMS Grid}
\tablehead{
\colhead{Input} & \colhead{Alignment Method} & \colhead{$M_1$} & \colhead{$M_2$} & \colhead{$P$}& \colhead{$\dot{M}_{transfer}$} & \colhead{$T_{eff,1}$} & \colhead{$T_{eff,2}$}& \colhead{$L_1$} & \colhead{$L_2$} & \colhead{$R_1$} & \colhead{$R_2$}}
\startdata
Change Points (4)   & -               & 1.8122 & 0.3381 & 3.2128 & 33.675  & 1.4933 & 0.3972 & 1.1133 & 0.4695 & 9.716  & 4.036  \\
Change Points (4)   & DTW (0-1 norm)  & 1.7594 & 0.332  & 3.1172 & 30.914  & 1.3727 & 0.4062 & 1.1257 & 0.4635 & 8.8358 & 3.966  \\
Change Points (4)   & DTW (max norm)  & 1.6931 & 0.3318 & 3.1442 & 31.3186 & 1.3926 & 0.3744 & 1.0952 & 0.457  & 8.7349 & 3.9666 \\
Change Points (4)   & CDTW (radius:3) & 1.7594 & 0.332  & 3.1172 & 30.914  & 1.3727 & 0.4062 & 1.1257 & 0.4635 & 8.8358 & 3.966  \\
Change Points (4)   & SDTW            & 1.7836 & 0.3339 & 3.2036 & 30.7565 & 1.4427 & 0.3835 & 1.0913 & 0.4654 & 8.9835 & 3.9371 \\
\hline
Change Points (8)   & -               & 0.9217 & 0.2486 & 1.8003 & 22.1917 & 0.9608 & 0.2417 & 0.5851 & 0.2386 & 8.0558 & 2.0624 \\
Change Points (8)   & DTW (0-1 norm)  & 0.7457 & 0.2373 & 1.5454 & 14.276  & 0.8898 & 0.2172 & 0.5184 & 0.1616 & 7.2545 & 1.853  \\
Change Points (8)   & DTW (max norm)  & 0.7175 & 0.2302 & 1.5892 & 16.8506 & 0.8679 & 0.2064 & 0.5015 & 0.1635 & 7.1702 & 1.79   \\
Change Points (8)   & CDTW (radius:3) & 0.7459 & 0.2372 & 1.5483 & 15.8113 & 0.8898 & 0.2172 & 0.5185 & 0.1632 & 7.2534 & 1.8536 \\
Change Points (8)   & SDTW            & 0.8581 & 0.2386 & 1.6742 & 18.091  & 0.9153 & 0.2183 & 0.5354 & 0.1757 & 7.4917 & 1.7885 \\
\hline
Change Points (16)  & -               & 0.596  & 0.2277 & 1.2938 & 23.2439 & 0.7345 & 0.1712 & 0.4104 & 0.1846 & 5.9181 & 1.5825 \\
Change Points (16)  & DTW (0-1 norm)  & 0.5465 & 0.2189 & 1.177  &  8.9193 & 0.5944 & 0.1246 & 0.2998 & 0.098  & 4.682  & 1.1629 \\
Change Points (16)  & DTW (max norm)  & 0.4964 & \textcolor{red}{0.2046} & 1.2259 & 13.0514 & 0.5621 & 0.1223 & 0.2858 & 0.0909 & 4.614  & 1.1778 \\
Change Points (16)  & CDTW (radius:3) & 0.5298 & 0.2173 & 1.1706 & 11.9898 & 0.5887 & 0.1247 & 0.3013 & 0.1004 & 4.6487 & 1.1641 \\
Change Points (16)  & SDTW            & 0.5237 & 0.2052 & 1.098  & 12.5501 & 0.619  & 0.1339 & 0.3121 & 0.107  & 4.9263 & 1.2269 \\
\hline
Change Points (32)  & -               & 0.5846 & 0.2327 & 1.2103 & 36.3708 & 0.6268 & 0.1464 & 0.3267 & 0.1192 & 4.8179 & 1.4481 \\
Change Points (32)  & DTW (0-1 norm)  & 0.5182 & 0.2197 & 1.1607 &  \textcolor{red}{8.3039} & 0.4316 & 0.1058 & 0.2121 & 0.0834 & 3.405  & 1.0446 \\
Change Points (32)  & DTW (max norm)  & \textcolor{red}{0.4541} & 0.209  & 1.2077 & 12.8097 & \textcolor{red}{0.3938} & 0.1036 & 0.2059 & 0.0796 & 3.2889 & 1.0235 \\
Change Points (32)  & CDTW (radius:3) & 0.4817 & 0.2169 & 1.1168 & 18.7697 & 0.4319 & 0.1111 & 0.2257 & 0.0833 & 3.4236 & 1.0797 \\
Change Points (32)  & SDTW            & 0.4891 & 0.2089 & \textcolor{red}{1.0608} & 13.0232 & 0.4314 & 0.109  & 0.2164 & 0.0734 & 3.544  & 1.1061 \\
\hline
Change Points (64)  & -               & 0.6882 & 0.2332 & 1.3281 & 39.5013 & 0.6803 & 0.1358 & 0.3285 & 0.1103 & 5.2436 & 1.3505 \\
Change Points (64)  & DTW (0-1 norm)  & 0.5389 & 0.2211 & 1.1902 &  8.401  & 0.4308 & 0.1034 & 0.2126 & 0.0885 & 3.4007 & 1.0487 \\
Change Points (64)  & DTW (max norm)  & 0.4643 & 0.2107 & 1.2416 & 13.5052 & 0.3945 & 0.1021 & \textcolor{red}{0.2043} & 0.0773 & \textcolor{red}{3.2719} & \textcolor{red}{1.0212} \\
Change Points (64)  & CDTW (radius:3) & 0.5387 & 0.2183 & 1.1558 & 26.0602 & 0.4882 & 0.1128 & 0.2444 & 0.0854 & 3.755  & 1.0917 \\
Change Points (64)  & SDTW            & 0.5089 & 0.2089 & 1.1228 & 15.0506 & 0.4648 & 0.109  & 0.2185 & \textcolor{red}{0.0696} & 3.6724 & 1.0811 \\
\hline
Change Points (128) & -               & 0.7585 & 0.247  & 1.4474 & 38.4068 & 0.7273 & 0.1462 & 0.3645 & 0.134  & 6.0697 & 1.492  \\
Change Points (128) & DTW (0-1 norm)  & 0.5461 & 0.2227 & 1.1925 &  8.3677 & 0.4298 & 0.1045 & 0.2173 & 0.0939 & 3.4154 & 1.0624 \\
Change Points (128) & DTW (max norm)  & 0.4688 & 0.2129 & 1.2708 & 13.143  & 0.3973 & \textcolor{red}{0.102}  & 0.2075 & 0.0805 & 3.2738 & 1.0489 \\
Change Points (128) & CDTW (radius:3) & 0.6177 & 0.2293 & 1.2774 & 28.65   & 0.5865 & 0.1236 & 0.2964 & 0.1075 & 4.8018 & 1.2261 \\
Change Points (128) & SDTW            & 0.5219 & 0.2109 & 1.1476 & 11.925  & 0.4679 & 0.1089 & 0.2316 & 0.0697 & 3.8267 & 1.0681 \\
\enddata
\tablecomments{Performance is measured as median relative absolute error in $\log_{10}$ scale. The first column indicates the number of change points used in each experiment. The second column shows the alignment method (DTW variants) applied in the ablation study, where '-' indicates no alignment. \textcolor{red}{Red} indicates the best results for each parameter. DTW (0-1 norm): change points are normalized between 0 and 1 before alignment. DTW (max norm): input values are normalized by dividing by the maximum signal value. CDTW (radius:3): Constrained DTW with radius parameter set to 3. Performance improvements are observed across all change point configurations, with the best results consistently achieved after applying DTW. This demonstrates that change points require proper alignment for optimal performance. Compared to our main results, the proposed method still outperforms the enhanced Change Point algorithm, indicating that superior performance stems from both better track alignment and the use of complete, unfiltered track data without.}
\end{deluxetable*}
%%%%%%%%%%%%%%%%%%%%%%%%%%%%%%%%%%%%%%%%%%%%%%%%%%%%%%%%%%%%%%%%%%%%%%%%%%%

\subsubsection{Alignment Parameter Selection}\label{sec:alignment_selection}
The alignment parameter subset $\mathcal{A}$ is selected empirically based on validation performance. For the HMS-HMS and CO-HeMS grids, using the full set of available parameters (as defined in Section \ref{sec:background}) for alignment produces the best interpolation accuracy. For the CO-HMS grid, using only the orbital period $P$ and luminosity $L_1$ for alignment yields better accuracy, likely because certain parameters in this grid exhibit sharp discontinuities that dominate the multi-dimensional distance computation when all parameters are included.

%%%%%%%%%%%%%%%%%%%%%%%%%%%%%%%%%%%%%%%%%%%%%%%%%%%%%%%%%%%%%%%%%%%%%%%%%%%

\subsection{Physical Consistency Analysis} \label{sec:physical_consistency}
While the error metrics and visual comparisons presented above evaluate interpolation accuracy for individual parameters, an important distinction must be made between morphological fidelity, which measures how well the interpolated tracks reproduce the shape of evolutionary sequences, and physical consistency, which concerns whether the predicted parameters collectively obey the physical laws governing stellar and binary evolution.

Our method is designed to preserve the morphological structure of evolutionary tracks during interpolation. This morphological preservation leads to improved physical consistency compared to the Change Point approach. In the initial formulation of our algorithm, each parameter was aligned independently, which could introduce temporal shifts between parameters of the same binary system. To address this, we extended the method to align all parameters simultaneously using a shared warping path, which enforces consistent temporal correspondence and preserves the ordering of major evolutionary features across all physical quantities.

Nevertheless, as a data-driven interpolation approach, our method does not explicitly enforce conservation laws, causal coupling between physical parameters, or consistency with the underlying differential equations governing stellar and binary evolution. The interpolated tracks should therefore be interpreted as data-driven surrogates rather than independently computed physical solutions. The method produces reliable results when the target initial conditions lie in the close vicinity of pre-simulated grid points, where the selected neighbors exhibit similar evolutionary behavior. In regions where neighbors are sparse or morphologically dissimilar, the alignment and averaging procedure may yield physically inconsistent predictions, as the method will still attempt to align tracks that lack meaningful correspondence.

Despite these limitations, we demonstrate below that our approach maintains key physical relationships in practice. We examine two complementary diagnostics: the coherence of coupled stellar parameters in the Hertzsprung-Russell diagram, and the preservation of the Stefan-Boltzmann relation across interpolated tracks.

%%%%%%%%%%%%%%%%%%%%%%%%%%%%%%%%%%%%%%%

\begin{figure*}
\includegraphics[width = \textwidth]{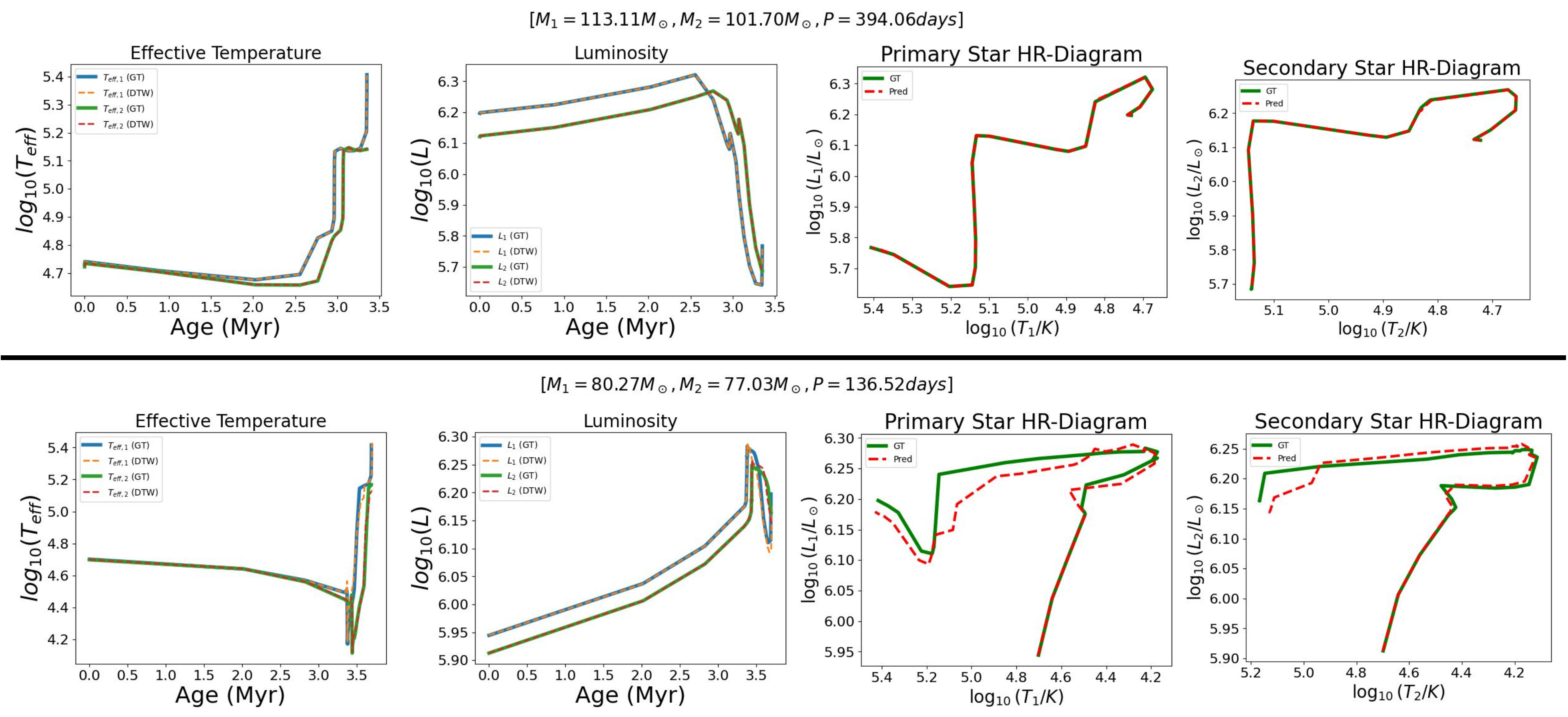}
\caption{Hertzsprung-Russell diagram analysis for two representative HMS-HMS systems. Each row corresponds to a different binary system with initial conditions shown above the panels. The first two panels display the effective temperature and luminosity as functions of age for both the primary and secondary stars, comparing ground truth (solid) and DTW-based predictions (dashed). The third and fourth panels show the corresponding HR diagrams for the primary and secondary stars, respectively. The top row presents a well-predicted system where the interpolated tracks closely follow the ground truth in both the time domain and the HR diagram. The bottom row shows a more challenging case where deviations occur during rapid evolutionary phases, resulting in visible differences in the HR diagram trajectories.
\label{fig:hr_diag}}
\end{figure*}

\subsubsection{Hertzsprung-Russell Diagram}

The HR diagram, which plots stellar luminosity against effective temperature, provides a direct visualization of whether coupled parameters remain physically coherent after interpolation. The HR diagram is one of the primary diagnostics through which stellar evolution is interpreted and is frequently used in comparison with observational counterparts such as color-magnitude diagrams. Since our method interpolates all parameters using a shared temporal alignment, the predicted $L$ and $T_{eff}$ values at each time step correspond to the same evolutionary state, preserving the expected trajectory through the HR diagram.

Figure \ref{fig:hr_diag} presents two examples from the HMS-HMS grid. The first system ($M_1 = 113.11\,M_\odot$, $M_2 = 101.70\,M_\odot$, $P = 394.06$\,days) demonstrates a case where the predicted effective temperatures, luminosities, and the resulting HR diagram trajectories closely follow the ground truth for both the primary and secondary stars throughout their evolution. The second system ($M_1 = 80.27\,M_\odot$, $M_2 = 77.03\,M_\odot$, $P = 136.52$\,days) illustrates a more challenging case where deviations are visible, particularly during the rapid evolutionary phases near the end of the tracks. The individual parameter predictions for $T_{eff}$ and $L$ show some discrepancies during these phases, which propagate into the HR diagram as deviations from the ground truth trajectory. Nevertheless, the overall morphology of the HR diagram tracks is preserved, and the predicted evolutionary path remains physically plausible.

%%%%%%%%%%%%%%%%%%%%%%%%%%%%%%%%%%%%%%%

\begin{figure*}
\centering
\includegraphics[width = 0.85\textwidth]{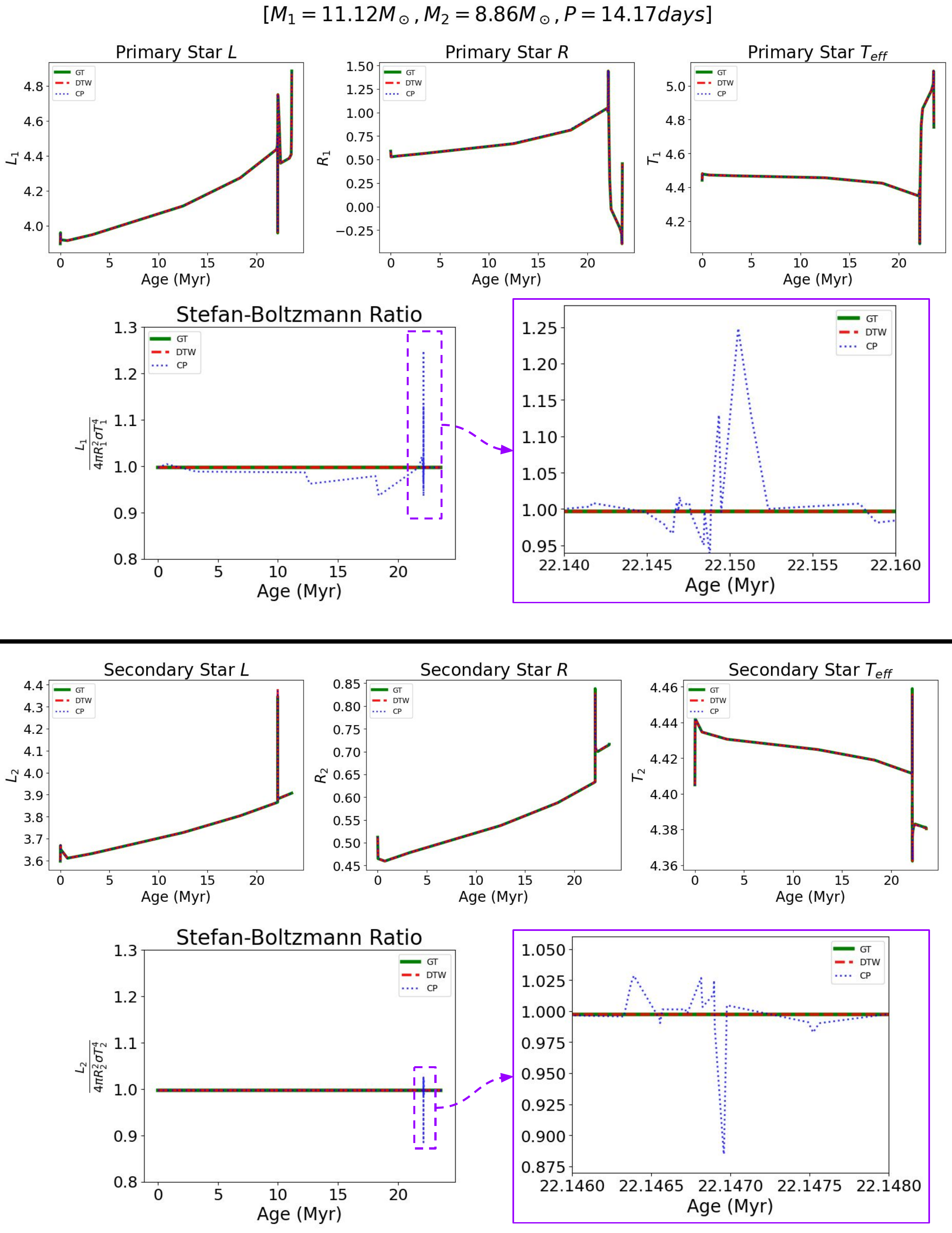}
\caption{Stefan-Boltzmann consistency analysis for a representative HMS-HMS system with initial conditions $M_1 = 11.12\,M_\odot$, $M_2 = 8.86\,M_\odot$, and $P = 14.17$\, days. The top half shows results for the primary star and the bottom half for the secondary star. For each star, the upper row displays luminosity, radius, and effective temperature as functions of age, comparing ground truth (solid green), DTW-based predictions (dashed red), and Change Point predictions (dotted blue). Below these, the Stefan-Boltzmann ratio $\mathcal{R}_{\mathrm{SB}} = L/(4\pi R^2 \sigma T_{\mathrm{eff}}^4)$ is shown for each method, with a zoomed-in view highlighting the rapid evolutionary phase. While the individual parameter predictions appear visually similar across all three methods, the Change Point approach produces spurious spikes in $\mathcal{R}_{\mathrm{SB}}$ during rapid evolutionary phases due to independent per-parameter sampling. The DTW-based joint alignment maintains $\mathcal{R}_{\mathrm{SB}} \approx 1$ throughout the evolution, consistent with the ground truth.}
\label{fig:sb_ratio}
\label{fig:sb_ratio}
\end{figure*}

\subsubsection{Stefan-Boltzmann Consistency}

The Stefan-Boltzmann law relates a star's luminosity $L$, radius $R$, and effective temperature $T_{\mathrm{eff}}$ through
\begin{equation}
L = 4\pi R^2 \sigma T_{\mathrm{eff}}^4,
\label{eq:stefan_boltzmann}
\end{equation}
where $\sigma$ is the Stefan-Boltzmann constant. For a physically self-consistent evolutionary track, the ratio
\begin{equation}
\mathcal{R}_{\mathrm{SB}} = \frac{L}{4\pi R^2 \sigma 
T_{\mathrm{eff}}^4}
\label{eq:sb_ratio}
\end{equation}
should remain close to unity at all time steps. Unlike the HR diagram, which examines the relationship between two parameters, $\mathcal{R}_{\mathrm{SB}}$ involves three quantities and provides a quantitative diagnostic for physical consistency.

Figure \ref{fig:sb_ratio} presents the luminosity, radius, and effective temperature predictions alongside the resulting Stefan-Boltzmann ratio for a representative HMS-HMS system ($M_1 = 11.12\,M_\odot$, $M_2 = 8.86\,M_\odot$, $P = 14.17$\,days), showing both the primary and secondary stars. The individual parameter predictions from the DTW-based method and the Change Point algorithm are visually indistinguishable from the ground truth across all three quantities. However, the Stefan-Boltzmann ratio reveals a critical difference between the two approaches. The ground truth and the DTW-based predictions both maintain $\mathcal{R}_{\mathrm{SB}} \approx 1$ throughout the evolution, while the Change Point predictions exhibit spurious spikes in the ratio, particularly during the rapid evolutionary phases near the end of the track. These spikes appear for both the primary and secondary stars despite the apparent accuracy of individual parameter predictions.

This discrepancy arises because the Change Point algorithm selects representative points independently for each parameter, producing a different age grid for each physical quantity. When the luminosity, radius, and temperature values are evaluated at a common set of time points, the temporal mismatches between parameters introduce artifacts in their combined relationships. In contrast, our joint-alignment approach places all parameters on a single consistent age grid, so the $L$, $R$, and $T_{eff}$ values at each time step correspond to the same evolutionary state. As a result, the Stefan-Boltzmann relation is preserved whenever the neighboring tracks used for interpolation satisfy it as well.

We emphasize that our method does not explicitly enforce the Stefan-Boltzmann law or any other physical constraint. The observed consistency is a consequence of the joint-alignment strategy, which preserves the temporal coupling between parameters that was present in the original simulated tracks.

We now show that if each neighboring track satisfies the Stefan-Boltzmann relation and all parameters are represented in logarithmic scale, the interpolated track preserves this relation exactly. The Stefan-Boltzmann law can be written in logarithmic form as
\begin{equation}
\log L = \log(4\pi\sigma) + 2\log R + 4\log T_{\mathrm{eff}}.
\label{eq:sb_log}
\end{equation}
This is a linear relationship among the logarithmic quantities. Let $\log L^k_t$, $\log R^k_t$, and $\log T^k_{\mathrm{eff},t}$ denote the aligned values of the $k$-th neighbor at time step $t$ after applying the shared warping path. If each neighbor satisfies Equation \ref{eq:sb_log} at every time step,
\begin{equation}
\log L^k_t = \log(4\pi\sigma) + 2\log R^k_t + 4\log 
T^k_{\mathrm{eff},t}, \quad \forall\, k, t,
\end{equation}
then the weighted average, with barycentric weights $b_k$ satisfying $\sum_{k=1}^{K} b_k = 1$, yields
\begin{align} 
\log L^*_t &= \sum_{k=1}^{K} b_k \log L^k_t \nonumber \\
&=  \sum_{k=1}^{K} b_k \Big[\log(4\pi\sigma) + 2\log R^k_t 
+ 4\log T^k_{\mathrm{eff},t}\Big] \nonumber \\
&=  \log(4\pi\sigma)\sum_{k=1}^{K} b_k + 2\sum_{k=1}^{K} b_k 
\log R^k_t + 4\sum_{k=1}^{K} b_k \log T^k_{\mathrm{eff},t} 
\nonumber \\
&=  \log(4\pi\sigma) + 2\log R^*_t + 4\log T^*_{\mathrm{eff},t},
\end{align}
where $\log R^*_t = \sum_k b_k \log R^k_t$ and $\log T^*_{\mathrm{eff},t} = \sum_k b_k \log T^k_{\mathrm{eff},t}$ are the interpolated values. Thus, the predicted track satisfies the Stefan-Boltzmann relation exactly at every time step.

This result relies on two conditions: (i) the parameters are represented in logarithmic scale, which transforms the nonlinear Stefan-Boltzmann law into a linear relation, and (ii) all parameters share a common temporal alignment through the joint warping path, so the values being averaged at each time step correspond to the same evolutionary state. If either condition is violated, the guarantee no longer holds. In particular, per-parameter alignment breaks condition (ii) by placing each quantity on a different age grid, which explains the spurious spikes observed in the Change Point results in Figure \ref{fig:sb_ratio}.

This formal guarantee, combined with the empirical results in Figure \ref{fig:sb_ratio}, confirms that joint alignment is essential for preserving physical relationships between coupled parameters in the interpolated tracks.

%%%%%%%%%%%%%%%%%%%%%%%%%%%%%%%%%%%%%%%%%%%%%%%%%%%%%%%%%%%%%%%%%%%%%%%%%%%
\section{Discussion}
\label{sec:discussion}

\subsection{Track Alignment and Interpolation}
Our results reveal that DTW effectively addresses the fundamental challenge of aligning irregularly sampled evolutionary tracks. DTW preserves all temporal information, while the Change Point algorithm reduces tracks to uniform length through selective sampling. Our approach proves that preserving the track morphologies is valuable during critical phases where rapid parameter changes require careful alignment to maintain prediction accuracy. Notably, when we apply DTW to tracks that have already been processed by the Change Point algorithm, DTW identifies alternative alignment patterns, demonstrating that change points are not optimally aligned across neighboring tracks.

The iterative weighted averaging framework successfully manages the complexity of aligning multiple neighbor tracks. By avoiding the consistency issues inherent in all-pairs alignment, our method maintains both computational efficiency and physical plausibility. 

\subsection{Comparison with Structure Preserving Schemes}
Structure-preserving interpolation schemes have been developed for stellar evolution within population synthesis frameworks such as the SEVN code \citep{Spera_2019, Iorio_2023}. These methods achieve physical fidelity by interpolating single-star evolutionary tracks using phase-specific coordinates, such as the percentage of stellar lifetime within each macro-phase (hydrogen burning, helium burning, carbon-oxygen core evolution). In these frameworks, binary systems are modeled by treating each stellar component independently: each star's properties are interpolated from a grid of single-star models, and when a binary interaction such as mass transfer occurs, the affected star is moved onto a different single-star track corresponding to its new mass. The binary interaction physics itself is handled through separate analytic prescriptions applied on top of the single-star interpolation. Because each individual track being interpolated follows standard single-star evolution, the phase boundaries remain well defined and serve as natural alignment anchors.

In contrast, MESA simulates both binary components simultaneously in a coupled calculation, where the interaction physics is self-consistently integrated into the evolutionary tracks. The resulting tracks capture how one star's mass loss directly affects the other star's internal structure in real time, producing complex, coupled evolutionary sequences that cannot be decomposed into independent single-star tracks. In these coupled binary tracks, mass transfer events can shift, interrupt, or entirely suppress standard evolutionary phases, making it impractical to define universal phase boundaries that apply across different binary configurations. Consequently, the phase-based alignment that underpins structure-preserving single-star interpolation methods cannot be directly applied to POSYDON's coupled binary evolution tracks. Our DTW-based approach addresses this gap by performing alignment in a data-driven manner that does not require predefined evolutionary markers, making it applicable to the complex and heterogeneous morphologies characteristic of coupled binary stellar evolution.

\subsection{Use Cases and Practical Considerations}
As discussed in Section~\ref{sec:physical_consistency}, the interpolated tracks should be interpreted as data-driven surrogates that inherit physical accuracy from the underlying MESA simulations and preserve it through the alignment and averaging procedure. The error distributions presented in Sections~\ref{sec:rae} and~\ref{sec:area_error} provide quantitative bounds on prediction accuracy, allowing users to assess the reliability of individual interpolated tracks for their specific application. Our method is well suited for statistical binary population synthesis, where individual track-level deviations are averaged out over large populations. For applications focused on individual systems, the reported error bars provide guidance on expected prediction uncertainty. The applicability of our method is currently demonstrated on POSYDON binary grids. The core algorithmic components, namely KNN retrieval, DTW alignment, and iterative weighted averaging, are general and do not depend on the specific physics implementation of POSYDON, but their effectiveness in a new context would need to be validated on the corresponding grid.

\subsection{Computational Requirements}
The computational efficiency gains achieved by our method are substantial. Full binary evolution simulations with MESA typically require in excess of $\sim$10 - 100 CPU hours per system \citep{fragos_posydon_2023}, representing a six-orders-of-magnitude increase in computational cost compared to rapid population synthesis codes. In contrast, our DTW-based interpolation method produces a complete predicted evolutionary track, including all 10 physical parameters, in approximately 0.01 seconds per binary system on a single core of an Intel Core i7-10700K CPU. At this rate, a population of $\sim$6,000 binaries can be interpolated in one minute, allowing the rapid generation of large binary populations that would otherwise require hundreds of CPU hours with full numerical simulations. The primary computational cost lies in the DTW alignment step, which roughly scales as $\mathcal{O}(X \times Y)$ for two tracks of lengths $X$ and $Y$. For typical track lengths in our grids (ranging from tens to thousands of time points), this computation remains tractable. Memory requirements are dominated by loading the pre-simulated grid for neighbor retrieval. The POSYDON grids require approximately 8-10 GB of memory \citep{fragos_posydon_2023, posydon_andrews}, which is well within the capabilities of standard workstations.

\subsection{Limitations and Future Work}
Despite clear improvements, several limitations merit discussion. Our reliance on K-nearest neighbors assumes that proximate grid points exhibit similar evolutionary behavior, an assumption that may fail in regions where binary evolution outcomes are highly sensitive to initial conditions. The effectiveness of barycentric weights depends on sufficient neighbor density, which decreases in sparse regions of the parameter space. Sparse grids can result in the selection of neighbors that exhibit significant morphological differences from one another. Even with proper temporal alignment, averaging dissimilar signals can deform the resulting interpolated track. However, selectively discarding neighbors based on morphological dissimilarity is non-trivial, as barycentric interpolation requires exactly four neighbors for our three-dimensional parameter space ($M_1$, $M_2$, $P$). Future work could explore alternative weighting schemes that accommodate variable numbers of neighbors or incorporate morphological similarity metrics into the neighbor selection process.

Beyond binary stellar evolution, our approach offers potential applications to other domains involving irregular time series interpolation where morphological preservation is critical. For example, climate scientists studying Earth's ancient climate face similar challenges when working with ice core data sampled at irregular time intervals with varying resolution. The combination of geometric neighbor selection with sequence alignment techniques provides a general framework for scientifically meaningful interpolation across multiple disciplines, including other areas of computational astrophysics and beyond. This versatility suggests that the methodological innovations presented here may find applications in diverse scientific contexts where similar challenges arise.

%%%%%%%%%%%%%%%%%%%%%%%%%%%%%%%%%%%%%%%%%%%%%%%%%%%%%%%%%%%%%%%%%%%%%%%%%%%
\section{Conclusion}
We presented a novel approach to binary stellar evolution track interpolation that addresses the fundamental challenges inherent in existing methodologies. By integrating DTW with barycentric weighted averaging, we have developed a framework that preserves the morphological integrity of evolutionary tracks while achieving substantial computational efficiency gains. A key extension of our method is the joint-alignment strategy, which computes a single shared warping path across all physical parameters simultaneously, placing them on a consistent temporal grid.

The experimental evaluation demonstrates consistent improvements across all evaluated parameters, with particularly striking enhancements for rapidly varying parameters such as mass transfer rates. The median relative absolute error reductions and area error improvements confirm that our approach generates more accurate predictions than existing change point algorithm \citep{Srivastava_2025}. Visual inspections further validate that interpolated tracks maintain the essential morphological features required for astrophysically meaningful analysis. Beyond morphological accuracy, we showed that the joint-alignment formulation preserves physical relationships between parameters. In particular, we proved that the Stefan-Boltzmann relation is exactly maintained in the interpolated tracks when the parameters are represented in logarithmic scale and the neighboring tracks satisfy the relation, a direct consequence of the linearity of weighted averaging in log space combined with the shared temporal alignment.

Our method is primarily intended for statistical binary population synthesis, where it enables the rapid generation of large populations at a fraction of the computational cost of full numerical simulations. The proposed method consistently outperforms existing approaches and provides a practical tool for efficient and accurate binary population synthesis studies.

%%%%%%%%%%%%%%%%%%%%%%%%%%%%%%%%%%%%%%%%%%%%%%%%%%%%%%%%%%%%%%%%%%%%%%%%%%%

\section{Acknowledgements}
We thank Tassos Fragos for fruitful relevant discussions. The POSYDON project is supported primarily by two sources: The Gordon and Betty Moore Foundation (PI Kalogera, grant awards GBMF8477 and GBMF12341) and a Swiss National Science  Foundation Professorship grant (PI Fragos, project No. CRSII5\_213497). The collaboration was also supported by the European Union’s Horizon 2020 research and innovation program under the Marie Sklodowska-Curie RISE action, grant agreement Nos. 691164 (ASTROSTAT) and 873089 (ASTROSTAT-II). Individual team members were supported by additional sources: J.J.A. acknowledges support for Program No.
JWST-AR-04369.001-A provided through a grant from the STScI under NASA contract NAS5-03127. M.B. was supported by the Boninchi Foundation, project No. CRSII5\_21349, and the Swiss Government Excellence Scholarship. V.K. was partially supported through the D. I. Linzer Distinguished University Professorship fund. U.D, P.M.S., A.K., V.K., S.G. and E.T. were supported by project Nos. GBMF8477 and GBMF12341. S.G., V.K., A.K., P.M.S. and U.D thank SkAI for support and hospitality for project operations. We gratefully acknowledge the support of the NSF-Simons AI-Institute for the Sky (SkAI) via grants NSF AST-2421845 and Simons Foundation MPS-AI-00010513.

%% For this sample we use BibTeX plus aasjournalv7.bst to generate the

\bibliography{main}{}
\bibliographystyle{aasjournalv7}



\end{document}